\documentclass[12pt,a4paper]{article}
\usepackage[utf8]{inputenc}
\usepackage{amsmath}
\usepackage{amsfonts}
\usepackage{amsthm}
\usepackage{amssymb}
\usepackage{mathrsfs}
\usepackage{color}
\usepackage{authblk}
\usepackage{enumerate}
\usepackage{xcolor}
\usepackage{rotating}
\usepackage[hidelinks]{hyperref}
\hypersetup{colorlinks,linkcolor={red!50!black},citecolor={blue!50!black},urlcolor={green!50!black}}
\usepackage[title]{appendix}
\theoremstyle{plain}
\newtheorem{theorem}{Theorem}
\newtheorem{lemma}[theorem]{Lemma}
\newtheorem{proposition}[theorem]{Proposition}
\newtheorem{corollary}[theorem]{Corollary}

\newtheorem{definition}[theorem]{Definition}
\usepackage{braket}
\newtheorem{conj}[theorem]{Conjecture}

\theoremstyle{plain}
\theoremstyle{remark}
\newtheorem{remark}[theorem]{Remark}
\theoremstyle{definition}
\newtheorem{example}[theorem]{Example}

\newcommand{\Fqmr}{\mathbb{F}_{q^{mr}}}
\newcommand{\CF}{\mathcal{F}}
\renewcommand{\l}{\lambda}
\newcommand{\Fq}{\mathbb{F}_q}
\newcommand{\Fqm}{\mathbb{F}_{q^m}}

\newcommand{\fql}{\phi_{q,\l}}
\newcommand{\K}{\mathbb{K}}
\newcommand{\F}{\mathbb{F}}
\newcommand{\N}{\mathbb{N}}
\newcommand{\R}{\mathbb{A}}
\newcommand{\D}{\mathcal{D}}
\newcommand{\B}{\mathcal{B}}
\newcommand{\uu}{\mu}
\newcommand{\f}{\phi}
\newcommand{\WW}{\mathbf{W}}
\renewcommand{\AA}{{A}}

\newcommand{\rank}{\textit{rank}}
\newcommand{\dr}{{d}_r}
\renewcommand{\aa}{\mathbf{a}}
\newcommand{\xx}{\mathbf{x}}
\newcommand{\yy}{\mathbf{y}}

\newcommand{\ff}{\mathbf{f}}
\newcommand{\g}{\mathbf{g}}

\newcommand{\cc}{\mathbf{c}}
\newcommand{\0}{\mathbf{0}}
\newcommand{\C}{\mathcal{C}}
\newcommand{\GG}{G}

\newcommand{\II}{{I}}
\newcommand{\MM}{{M}}

\newcommand{\PP}{{P}}
\newcommand{\E}{\mathbb{E}}
\newcommand{\n}{\eta}
\newcommand{\Ev}{\mathbf{Ev}}
\newcommand{\Lf}{\mathcal{L}[X;\phi]}
\newcommand{\G}{\mathcal{G}}
\renewcommand{\a}{\alpha}
\newcommand{\sr}{\textit{sumrank}\,}
\newcommand{\<}{\left<}
\renewcommand{\>}{\right>}

\author[1]{Rakhi Pratihar\thanks{Email: rakhi.pratihar@uit.no- During the course of this work, Rakhi Pratihar was  supported by a doctoral fellowship at IIT Bombay from the University Grant Commission, Govt. of India (Sr. No. 2061641156). Currently, she is supported by Grant 280731 from the Research Council of Norway.}}
\author[2]{Tovohery Hajatiana Randrianarisoa \thanks{Email:(Corresponding author) tovo@aims.ac.za - The author was supported by the Swiss SNF grant No. 181446 while he was working at the Indian Institute of Technology Bombay}}
\affil[1]{Department of Mathematics and Statistics, %\newline \indent
		UiT - The Arctic University of Norway}
\affil[2]{Department of Mathematical Sciences, 
Florida Atlantic University }

\title{Constructions of optimal rank-metric codes from automorphisms of rational function fields}
\begin{document}
\maketitle

%\begin{abstract}
%We define a class of automorphisms of fields of rational functions over finite fields and discuss its applications in various constructions of linear codes. The main construction is of maximum rank distance (MRD) codes over rational function fields of finite characteristic. Reducing these codes to finite fields, we obtain a family of MRD codes which are not equivalent to generalized twisted Gabidulin codes under certain conditions. We also consider rank metric codes in Ferrers diagrams which are being used in multilevel construction of subspace codes. This paper contributes in proving further cases of a conjecture by Etzion and Silberstein on existence of optimal Ferrers diagram rank metric codes. The new construction produces optimal codes for some Ferrers diagrams and parameters for which no constructions of optimal codes were known before. We also show that the codes over rational function fields can be applied to recover some existing maximum sum rank distance codes which generalize both Gabidulin and Reed-Solomon codes. 

%\textbf{Keywords}: Rank metric, Sum rank metric, Ferrers diagram, Optimal linear codes, Generalized twisted Gabidulin codes, Automorphisms, Rational functions, Finite fields. 

%\textbf{MSC}: 15A03, 94B05, 94B60.
%\end{abstract}

%%%%%%%%%%%%%%%%%%%%%%%%%%%%%%%%%%%%%%% Section 1 %%%%%%%%%%%%%%

\begin{abstract}
We define a class of automorphisms of rational function fields of finite characteristic and employ these to construct different types of optimal linear rank-metric codes. The first construction is of generalized Gabidulin codes over rational function fields. Reducing these codes over finite fields, we obtain maximum rank distance (MRD) codes which are not equivalent to generalized twisted Gabidulin codes. We also construct optimal Ferrers diagram rank-metric codes which settles further a conjecture by Etzion and Silberstein. 
%We also recover some existing maximum sum rank distance codes using the construction over rational function fields.
%In particular, these MRD codes over finite fields induce linear complementary pair (LCP) of codes. 
%Based on the MRD codes over rational functions, we construct Ferrers diagram rank-metric codes which achieve the largest possible dimension and thus consolidating further a conjecture by Etzion and Silberstein. 

\textbf{Keywords}: Automorphisms, Rational functions, Finite fields, rank-metric, Sum rank-metric, Ferrers diagram, Optimal linear codes, Generalized twisted Gabidulin codes. 

\textbf{MSC}: 15A03, 94B05, 94B60.
\end{abstract}

%%%%%%%%%%%%%%%%%%%%%%%%%%%%%%%%%%%%%%intro%%%%%%%%%%55
\section{Introduction}\label{sec:1}
Rank-metric codes are $q$-analogues of classical codes with Hamming metric. These codes were introduced independently by Delsarte \cite{Del78} in 1978 and Gabidulin \cite{Gab85} in 1985 from a combinatorial point of view. Their application to crisscross error correction was obtained by Roth \cite{Rot91} in 1991. Where cryptography is concerned, Gabidulin, Paramonov and Tretjakov proposed the GPT cryptosystem in \cite{GPT91}, which is a modification of the McEliece cryptosystem \cite{McE78}, by replacing the binary Goppa codes with rank-metric codes. In 2008, Silva, K\"oetter and Kschischang adapted rank-metric codes for error-correction in random network coding \cite{SKK08}. 

For a prime power $q$ and positive integers $m,n$, Delsarte introduced a rank-metric code as a $\Fq$-subspace of the matrix space $\Fq^{m \times n}$, while Gabidulin considered it as a $\Fqm$-linear subspace of $\Fqm^n$. By fixing an $\Fq$-basis of $\Fqm$, it is easy to see that these two representations are equivalent as a vector in $\Fqm^n$ can be represented as an $m \times n$ matrix over $\Fq$. The \emph{rank distance} of two codewords is the usual matrix rank of their difference for Delsarte rank-metric codes or difference of their matrix representations for Gabidulin rank-metric codes. The minimum of rank distances between any two distinct codewords is called \emph{minimum rank distance} of the code. For Gabidulin rank-metric codes it is proved in \cite{Del78} that the maximum possible dimension of such a code with minimum rank distance $d$ is $n-k+1$ and the codes attaining this bound is called \emph{maximum rank distance} (MRD) codes.  
The first known class of MRD codes over finite fields for any admissible parameters was independently constructed by Delsarte \cite{Del78}, Gabidulin \cite{Gab85}, and Roth \cite{Rot91}. Although the first construction was given by Delsarte, these codes are known as \emph{Gabidulin codes} in the literature. These codes can be expressed using $q$-polynomials or linearized polynomials, which was introduced and studied by Ore \cite{Ore331,Ore332}. 
 There are cryptosystems, other than GPT, which partially use Gabidulin codes \cite{RQC,ABDGZ18}. But there are several constructions of non-Gabidulin MRD codes also, for example, see \cite{HTM17,KG05,OO16,She16,She18}. In fact, it is theoretically proved in \cite{NHTRR18} that there exist non-Gabidulin MRD codes when the field size is large enough.

More recently, Augot \textit{et al.} have extended the notion of rank metric to arbitrary field extensions, possibly of characteristic zero in \cite{ALR18}. They have used $\theta$-polynomials \cite{Ore331} which generalizes linearized polynomials and named the codes as generalized Gabidulin codes. In \cite{ALR18}, the authors studied particularly the  generalized Gabidulin codes over integer rings and their quotient fields. They proved that the reduction of these codes over finite fields gives back the original Gabidulin codes over finite fields. On the other hand, in \cite{Ove08} Overbeck proved that the GPT cryptosystem (and its variants) is insecure due to the algebraic structure of the (Gabidulin) codes. Therefore, it is natural to ask whether there exists generalized Gabidulin code whose reduction over finite fields is an MRD code which is not equivalent to any generalized twisted Gabidulin code. We answer this question affirmatively. 

 We give a particular construction (Proposition \ref{PRO25}) of generalized Gabidulin codes over $\Fqm(x)$, i.e. rational function fields in one variable over finite fields by considering a suitable class of automorphisms of $\Fqm(x)$ and then considering the extension $\Fqm(x)/\K$ where $\K$ is the subfield of $\Fqm(x)$ fixed under one such automorphism (Proposition \ref{pro:21} and Theorem \ref{thm:24}). We also give constructions of a large class of MRD codes over finite fields (Theorem \ref{thm:28}) which are not equivalent to the generalized twisted Gabidulin codes introduced in \cite{She16}. We obtain these codes by reducing the MRD codes over rational function fields modulo an irreducible polynomial of some preferred degree (Theorem \ref{thm:26}).

Another application of the newly constructed MRD codes over rational functions that we discuss in this paper is the construction of Ferrers diagram rank-metric codes. 
In \cite{SKK08}, Silva \textit{et al.} introduced a method of constructing subspace codes by lifting linear matrix rank-metric codes in $\Fq^{m\times n}$. But even MRD codes may not give subspace codes of largest possible size after lifting. To improve the result Etzion and Silberstein proposed a multilevel approach in \cite{ES09} using Ferrers diagram rank-metric codes. Such codes consist of matrices having zeroes at certain fixed positions determined by the Ferrers diagram (see Definition \ref{defn:30}). In the same paper, the authors have formulated a bound on the dimension of Ferrers diagram rank-metric codes and the codes attaining that bound are called \textit{optimal}. In order to get a subspace code with size as large as possible, we can use optimal Ferrers diagram rank-metric codes. It is also conjectured in \cite{ES09} that optimal Ferrers diagram rank-metric codes exist for any set of parameters. Several constructions of optimal Ferrers diagram rank-metric codes can be found in \cite{AG19,EGRWZ16,ES09,LCF19,ZG19}, still there are many Ferrers diagrams for which no construction is known for optimal codes. In this paper, we show how some subcodes of the MRD codes over rational function fields provide optimal rank-metric codes for some Ferrers diagrams (Theorem \ref{thm:37} and Proposition \ref{pro:42}). 

%A multilevel construction is proposed in \cite{NU09,NU10} and the approach is an improvement compared to the use of concatenations of one-shot rank-metric codes. The use of convolutional rank-metric codes for application in multishot network coding were also studied \cite{MBK15,NPRV17,WS12,WSB12,WSS15} and a survey on convolutional codes can be found in \cite{NS18}.
 The previously mentioned network coding is called one-shot network coding as they use the network channel only once. When the network channel is used multiple times, it is called multishot network coding. For this type of network coding, the sum rank-metric codes were proposed in \cite{MBK15,MP19,MK19,MK18,NU10}. These codes can also be used in space time coding \cite{GH03} and in distributed storage \cite{MK18}. In the Appendix, we provide a new way of constructing the maximum sum rank distance (MSRD) codes of \cite{MK19} by using our construction of MRD codes over rational function fields (cf. Definition \ref{def:50} and Theorem \ref{thm:51}).

  The paper is organized as follows. In the next section, we collect some preliminaries about rank-metric codes over finite fields. In Section \ref{sec:3} we recall the definitions and basic notions regarding the generalized Gabidulin codes over arbitrary algebraic extensions from \cite{ALR18} and we give a characterization of MRD codes over arbitrary fields.
In Section \ref{sec:4}, we define a class of automorphisms for the rational function fields with finite characteristic and study the extension over the fixed field under such an automorphism. Using these automorphisms, we construct generalized Gabidulin codes over rational functions in Section \ref{sec:5} and show when their reduction over finite fields gives MRD codes which are not equivalent to the generalized twisted Gabidulin codes. Section \ref{sec:6} deals with constructions of optimal Ferrers diagram rank-metric codes based on the MRD codes over rational functions. 
In the appendix, we show an alternative way of constructing the linearized Reed-Solomon codes of \cite{MK19}, a class of MSRD codes, using the automorphisms of rational functions. 
%%%%%%%%%%%%%%%%%%%%%%%%%%%%%%%%%%%%%%%%%%%%%%%%%%%%%%%%%%%%%%%%555
%%%%%%%%%%%%%%%%%%%
\section{Preliminaries}\label{sec:2}
Throughout, $\Fqm$ denotes the finite extension of degree $m$ over the finite field $\Fq$ for some prime power $q$ and positive integer $m$. For positive integers $m,n$ and a field $\F$, we use $\F^{m \times n}$ to denote the space of all $(m \times n)$ matrices with entries in $\F$. Given any $c_1, \ldots, c_n \in \Fqm$, we denote by $\<c_1,\dots,c_n\>_{\Fq}$ the $\Fq$-linear subspace of $\Fqm$ generated by $c_1, \ldots, c_n$.
 
 %We briefly review the basic theory of rank-metric codes and the results which will be used in the subsequent sections.
 The following definition of rank-metric codes is by Gabidulin \cite{Gab85} which is essentially same as the definitions by Delsarte \cite{Del78} and Roth \cite{Rot91}.

%rank-metric codes was introduced by Delsarte in \cite{Del78} and later rediscovered by Gabidulin \cite{Gab85} and Roth \cite{Rot91} independently. Here we give the definition by Gabidulin in which is essentially the 
\begin{definition}
For a positive integer $n$, the \emph{rank} of an element $\cc = (c_1,\dots,c_n)$ in  $\Fqm^n$ is defined by 
\[
rank(\cc) := \dim\<c_1,\dots,c_n\>_{\Fq}.
\]
The rank function induces a metric $\dr$, called \emph{rank metric}, on
$\Fqm^n$ where $\dr(\cc,\cc') := rank(\cc - \cc')$ for $\cc,\cc'$ in $\Fqm^n$.  

A (linear) rank-metric code $\C$ over the finite extension $\Fqm/\Fq$ of length $n$ and dimension $k$ is an $\Fqm$-subspace of $\Fqm^n$ of dimension $k$ endowed with the induced rank metric $\dr$. If $d = \min\{rank(\cc): \cc \in \C\backslash \{\0\}\}$, then we call $\C$ an $[n,k,d]_{\Fqm/\Fq}$-rank-metric code.
 \end{definition}

When there is no ambiguity about the fields, we use the notation $[n,k,d]$ to denote a rank-metric code of length $n$ with dimension $k$ and minimum rank distance $d$. 

%For more on the theory of linear rank metric codes over finite fields, one can see \cite{Gab85,Del78}.

Analogous to the classical case, rank-metric codes satisfy the following Singleton-like bound.
\begin{proposition}\cite{Del78}\label{pro:2}
Let $\C$ be a $[n,k,d]$-linear rank-metric code over $\Fqm/\Fq$, then $d\leq n-k+1$. 
\end{proposition}

Codes attaining the Singleton bound are called maximum rank distance (MRD) codes.
The very first construction of MRD codes over finite fields for any admissible parameters was independently given by Delsarte \cite{Del78}, Gabidulin \cite{Gab85}, and Roth \cite{Rot91}. These MRD codes are widely known as \emph{Gabidulin codes} and are defined as follows.

\begin{definition}\label{exa:2}
Let $m, n$ and $k$ be positive integers such that $k \leq n \leq m$. Suppose $\aa = (a_1, \dots, a_n ) \in \Fqm^n$ is such that the $a_i$'s are $\Fq$-linearly independent. Then the Gabidulin code over $\Fqm/\Fq$ of length $n$ and dimension $k$ is defined as the code with $G$ as a generator matrix, where
\[
	G = 
	\begin{pmatrix}
	a_1 & a_2 & \hdots & a_n \\
	{a_1}^q & {a_2}^q & \hdots & {a_n}^q \\
	\vdots & \vdots & \ddots & \vdots \\
	{a_1}^{q^{k-1}} & {a_2}^{q^{k-1}} & \hdots & {a_n}^{q^{k-1}} \\
	\end{pmatrix}.
	\]
\end{definition}

These codes can be alternatively represented as $q$-polynomials or linearized polynomials over $\Fqm$. Here we briefly review the correspondence of the two representations. 
\begin{definition}
    A linearized polynomial over $\Fqm$ is of the form $\sum_i f_i X^{q^i}$
 where $f_i \in \Fqm$ and only finitely many $f_i$'s are nonzero. The {\em $q$-degree} of this linearized polynomial is the largest $i$ such that $f_i$ is nonzero. By convention, $q$-degree of the zero polynomial is assumed to be $-\infty$.
 
 We use $\mathcal{L}_k[X]$ to denote the set of all linearized polynomial over $\Fqm$ of $q$-degree at most $k-1$, i.e.,  
 \[
   \mathcal{L}_k[X] := \{f_0X + f_1X^q + \ldots+ f_{k-1}X^{q^{k-1}}  \colon f_i \in \Fqm\}.
\]  
\end{definition}
 For the theory of linearized polynomials, one can refer \cite{Ore331,Ore332} by Ore.
 
 Now it is clear that the Gabidulin code $
\mathcal{G}_{k}(\aa) := \{ \left(f(a_1), \dots, f(a_n) \right)\colon f(X)\in \mathcal{L}_k[X]\}.
$

%The above construction was independently invented in \cite{Del78,Gab85,Rot91}. The codes are known as Gabidulin codes 

These codes were further generalized in \cite{KG05} by replacing $x^q$ with $x^{q^s}$ where $s$ is an integer such that $\gcd(m,s)=1$. In this case, the codes are called \emph{generalized Gabidulin codes}. %and denoted by $\mathcal{G}_{k,s}(\aa)$.%
These codes admit fast decoding algorithm \cite{Gab85}. In \cite{She16}, Sheekey gives a new family of linear MRD codes which strictly contains the already known classes of MRD codes, i.e. (generalized) Gabidulin codes. These codes are known as \emph{twisted Gabidulin codes}. Sheekey's construction \cite{She16} is a generalization of the construction of Otal \emph{et al.} in \cite{OO16}.

Let $k \leq n \leq m$ be integers. Then for $\n \in \Fqm$, consider the following set of linearized polynomials over $\Fqm$:
 \[
   \mathcal{L}_k[X;\eta,h] = \{f_0x + f_1x^q + \ldots+ f_{k-1}x^{q^{k-1}} + \n f_0^{q^h}x^{q^k} \colon f_i \in \Fqm\}.
\]  
Originally, twisted Gabidulin codes were defined as linearized polynomials $\mathcal{L}_k[X;\eta,h]$. When seen as $\Fq$-linear operators of $\Fqm$, they define matrix rank-metric codes in $\Fq^{m \times n}$ of dimension $mk$.  
Note that $\mathcal{L}_k[X;\eta,h]$ is $\Fqm$-linear space if and only if $h = 0$. In the following definition, we consider only those twisted Gabidulin codes which are $\Fqm$-linear.  

\begin{definition}\label{exa:1}
Let $\Fqm/\Fq$ be a finite field extension of degree $m$ and let $m, n, k $ be positive integers such that $k \leq n \leq m$. Let $\n \in \Fqm^\times$ such that $N(\n) \neq (-1)^{nk}$. Let $a_1, \ldots, a_n$ are $\Fq$-linear independent elements of $\Fqm$. Then the twisted Gabidulin code over $\Fqm/\Fq$ of length $n$ and dimension $k$ is defined as
\[
\mathcal{H}_k(\eta,h=0) := \{ \left(f(a_1), \dots, f(a_n) \right)\colon f(X)\in \mathcal{L}_k[X;\eta,h=0] \}.
\]
\end{definition}

 Twisted Gabidulin codes are generalizations of the Gabidulin codes in the sense that we get back the later when $\n=0$. Similar to Gabidulin codes, the codes in Definition \ref{exa:1} can also be generalized by replacing the Frobenius map $x \mapsto x^{q}$ with an automorphism $x \mapsto x^{q^s}$ \cite{She16}. These codes, known as generalized twisted Gabidulin codes, were studied in \cite{LTZ18}.

There are also several constructions of rank-metric codes which are not MRD, see for example \cite{AGHRZ19}. A survey on rank-metric codes over finite fields can be found in \cite{She19}.

Next we recall a characterization of generalized twisted Gabidulin codes that will be used later.  

\begin{proposition}[\cite{GZ19,HTM17}]\label{pro:3}
Let $\C \subseteq \Fqm^n $ be a linear MRD code of dimension $k < n$. If $\C$ is a generalized twisted Gabidulin code, then $\dim_{\Fqm} \C \cap \C^{q^s} \geq k-2$ for some integer $s$ with $\gcd(m,s)=1$, where
\[
 \C^{q^s} = \{(c_1^{q^s}, \dots, c_n^{q^s}) \colon (c_1, \dots, c_n) \in\C \}.
 \]	
\end{proposition}

Proposition \ref{pro:3} only provides a necessary condition for an MRD code to be a generalized twisted Gabidulin code. However, when the dimension $\dim \C \cap \C^{q^s}$ is equal to $k-1$ then the converse is also true for generalized Gabidulin codes \cite[Theorem 4.8]{HTM17}.

%Later we want to construct codes which are not equivalent to the generalized twisted Gabidulin codes. For that we first want to define
There are different notions of equivalence of rank-metric codes \cite{Ber03,Mor14}. Here we consider the one from \cite{Mor14} based on $\Fqm$-linear isometries.

\begin{definition}
Two linear rank-metric codes $\C_1$ and $\C_2$ of length $n$ over $\Fqm/\Fq$ are equivalent if there exist $\alpha \in \Fqm^\times$ and an invertible matrix $\MM\in\Fq^{n\times n}$ such that $\C_1 = \alpha \C_2\MM$, where
\[
\alpha \C_2\MM := \{(\alpha c_1,\dots,\alpha c_n)\MM\colon (c_1,\dots,c_n)\in \C_2)\}.
\]
\end{definition}

The above definition implies that any code equivalent to a generalized twisted Gabidulin code is also a generalized twisted Gabidulin code.

\section{Generalization of rank metric over arbitrary fields}\label{sec:3}

%Recently rank metric codes over arbitrary field (not necessarily finite) extension of finite degree have also been considered. Roth mentioned these codes in \cite{Rot96} and further studies on rank metric codes over general fields are presented in \cite{Aug14,ALR18}. Here we recall some relevant results of rank metric code over general fields from \cite{ALR18}. We also generalize some properties of rank metric codes over finite fields to those over general fields. 

Throughout this section, $\F, \E$ denote arbitrary fields such that $\F\hookrightarrow \E$ is an extension of finite degree $m$.
We use $Aut_{\E}(\F)$ to denote the group of automorphisms of $\F$ fixing $\E$.

We recall the notion of rank metric for arbitrary algebraic extension as introduced in \cite{Rot96} and further in \cite{ALR18}. Among the four equivalent definitions (see, \cite[Definition 8]{ALR18}), we record the one which is compatible with the finite fields case.
\begin{definition}\label{def:2}
Let $n, m$ be positive integers and $\F/\E$ be an extension of degree $m$. For an element $\xx= (x_1,\dots,x_n) \in \F^n$, its rank, denoted as $\rank(\xx)$, is equal to the dimension of the $\E$-subspace of $\F$ generated by $x_i$'s.
The rank function induces a metric $\dr$, called \emph{rank metric} on $\F^n$ where $\dr(\xx, \yy) := \rank (\xx-\yy)$ for $\xx, \yy \in \F^n$.

A rank-metric code $\C$ over the finite extension $\F/\E$ of length $n$ and dimension $k$ is a $k$-dimensional $\F$-subspace of $\F^n$ endowed with the rank metric $\dr$. If $d = \min\{rank(\cc): \cc \in \C \setminus \{\0\}\}$ we call $\C$ to be an $[n,k,d]$-linear rank-metric code over $\F/\E$.
\end{definition}

In \cite{ALR18}, it is proved that the $[n,k,d]$ rank-metric codes over arbitrary algebraic extensions also satisfy the Singleton bound $d \leq n-k+1$ similar to the case of finite fields. If equality holds, then the codes are called MRD codes.

The next result provides a necessary and sufficient condition for a rank-metric code to attain the Singleton bound. This is an extension of the finite fields case as proved in \cite{HTM17}. 

\begin{theorem}\label{thm:3}
Let $\E \hookrightarrow \F$ be a finite extension of arbitrary fields such that $[\F \colon \E] =m$. Let $k\leq n\leq m$ and let $\GG\in \F^{k\times n}$ be a generator matrix of an $[n,k]$-rank-metric code $\C$ over $\F/\E$. Then $\C$ is an MRD code if and only if $\GG\MM$ is invertible for any $\MM \in \E^{n\times k}$ of rank $k$.
\end{theorem}
\begin{proof}
Suppose $\C$ is an MRD code and let $\MM \in \E^{n\times k}$ of rank $k$ be such that $\GG\MM$ is not invertible. So there is $\xx \in \F^k\setminus \{\0\}$ such that $\xx\GG\MM=\0$. Now the rank of the codeword $\cc=\xx\GG$ of $\C$ is equal to the rank of $\phi_\cc $ as an $\E$-linear map $\phi_\cc\colon\E^{n} \longrightarrow \F$ defined as dot product with $\cc$. Thus rank-nullity theorem implies that $\rank(\cc)$ is at most $n-k$, which contradicts our assumption of $\C$ being MRD. This shows that $\GG\MM$ is invertible for any $\MM \in \E^{n\times k}$ of rank $k$.

Conversely, let $\GG\MM$ be invertible for any matrix $\MM \in \E^{n\times k}$ of rank $k$. Suppose $\rank(\xx\GG) \leq n-k$ for some $\xx \in \F^k\setminus \{\0\}$. 
Without loss of generality, we suppose that $\xx\GG=(c_1,\ldots,c_{n-k},\ldots,c_n)$ where the $c_i$'s are $\E$-linear combinations of $(c_1,\ldots,c_{n-k})$ for $n-k< i\leq n$. 
Then consider $\AA$ to be the $((n-k)\times k)$-matrix whose $i$-th column is a coefficient matrix of $c_i$ when written as linear combination of $\{c_1,\ldots,c_{n-k}\}$. But then for the matrix $\MM = \begin{bmatrix} \AA \\ -\II_k\end{bmatrix}$ of rank $k$ we have $\xx\GG\MM=\0$.
This contradicts the injectivity of $\GG\MM$ and it completes the proof.
\end{proof}

%In the rest of this section, we recall the construction of MRD codes over arbitrary algebraic extension $\E \hookrightarrow \F$ of finite degree by Augot et al. in \cite{ALR18}. The construction generalize the (generalized) Gabidulin codes over arbitrary fields by replacing the linearized polynomials with their natural generalization called \emph{$\phi$-polynomials}, where $\phi\in Aut_\E(\F)$. 

In \cite{ALR18}, the authors have introduced $\phi$-polynomials for $\phi\in Aut_\E(\F)$ to extend the generalized Gabidulin codes over arbitrary fields.
\begin{definition}\cite[Definition 1]{ALR18}
 A $\phi$-polynomial over $\F$ is a finite sum of the form $\sum_{i \geq 0} f_iX^i, ~ f_i \in \F.$ The largest integer $i$ such that $f_i \neq 0$ is called the $\phi$-degree of the polynomial. By convention, the degree of the zero polynomial is $- \infty$.
\end{definition}
We denote the set of $\phi$-polynomials by $\Lf$ and the subset of $\phi$-polynomials of $\phi$-degree at most $k-1$ by $\mathcal{L}_k[X;\phi]$.

The set $\Lf$ is a non-commutative algebra over $\F$ with component-wise addition and the symbolic product: $$ \sum_i f_i X^i \cdot \sum_i g_i X^i = \phi^i(g_j)X^{i+j}.$$

%-- (component-wise addition): $L_1(X) + L_2(X) = \sum_i (f_i + g_i) X^i$;

%-- (\emph{symbolic} product): $L_1(X) \cdot L_2(X) = \sum_{i,j} f_i \phi^i(g_j)X^{i+j}$. 
 
% Unlike in the case of usual polynomial ring $\F[X]$, the evaluation of a $\phi$-polynomial $L(X) = \sum_i f_i X^i$ at an element $a \in \F$ is defined by 
% \[ L(a) = \sum_i f_i \phi^i(a).\]

%It is well known that the ring of linearized polynomials over $\Fqm$ is isomorphic to the ring of matrices $\Fq^{m \times m}$.

There is a well-known bijection between the set of linearized polynomials over $\Fqm$ and the set of $\Fq$-linear operators of $\Fqm$. Similarly, we can consider $\phi$-polynomials over $\F$ as an $\E$-linear operator of $\F$ as follows;
\[L(X) = \sum_i f_i X^i \mapsto L_{\phi} = \sum_i f_i \phi^i: a \mapsto \sum_i f_i \phi^i(a).\]
%({\color{green}Do we need to write this here?})Note that the linear operator $L_\phi$ is $\K$-linear where $\K$ is the fixed field of $\phi$, i.e., 
%\[
%\K = \{ f\in \F\colon \phi(f) = f \}.
%\]
%For any $f_i\in \F$, we define the linear operator
%\[
%L = f_0 + f_1 \phi + f_2 \phi^2 + \ldots + f_k \phi^k.
%\]

%These linear operators are evaluation of $\phi$-polynomials (see \cite[Definitions 1, 3]{ALR18}) which are generalization of linearized polynomials.
%\begin{definition}
%Let $\F/\E$ be a field extension and let $\phi\in Aut_\E(\F)$. The fixed field of $\phi$ is the set 
%\[
%\K = \{ f\in \F\colon \phi(f) = f \}.
%\]
%\end{definition}
%It is easy see that the fixed field is indeed a field. 
%
%The linear operator $L$ induces a $\K$-linear map  $\F \rightarrow \F$ by
%\[
%L(g) := f_0 g + f_1 \phi(g) + f_2 \phi^2(g) + \ldots + f_k \phi^k(g).
%\]
%For the rest of this chapter we will use the notation $\Lf$ to denote the set of $\phi$-polynomials and also the set of linear operators interchangeably and $\mathcal{L}_k[X;\phi]$ to denote those with $\phi$-degree at most $k-1$.
%\begin{definition}
%With the above notation, given a linear operator $L\in\Lf$, the kernel $\ker L$ of $L$ is the $\K$-vector space given by the set of elements $g\in \F$ such that $L(g) = 0$.
%\end{definition}

%Next we extend some well known facts in finite fields to the case of arbitrary fields.
To construct linear MRD codes over the extension $\F/\E$ we consider the following matrix.
%Later in this section we will see that the construction is same as the construction of generalized Gabidulin codes over arbitrary fields as introduced by Augot et al. in \cite{ALR18}.
%The difference between these two approaches are given in Remark \ref{rem:A}.
\begin{definition}
	Let $\phi \in Aut_\E(\F)$ and let $\aa = \{a_1,\ldots,a_n\}$ be a set of $n$ distinct elements  of $\F$. The $n$-th order Moore matrix  with respect to $\phi$ and $\aa$ is
	\[
	\WW_n(\aa,\phi) := 
	\begin{pmatrix}
	a_1 & a_2 & \hdots & a_n \\
	\phi(a_1) & \phi(a_2) & \hdots & \phi(a_n) \\
	\vdots & \vdots & \ddots & \vdots \\
	\phi^{n-1}(a_1) & \phi^{n-1}(a_2) & \hdots & \phi^{n-1}(a_n) \\
	\end{pmatrix}.
	\]
\end{definition}

The following result shows the relation between the Moore matrix and the $\E$-linear independence of $a_1,\dots, a_n$.

\begin{proposition}\cite[Theorem 4]{ALR18}\label{PRO15}
Let $\F/\E$ be a field extension and $\phi\in Aut_\E(\F)$ with $\E$ being the fixed field of $\phi$. Let $\aa = \{a_1,\dots,a_n\}$ be a set of $n$ distinct elements of $\F$. Then the $a_i$'s are linearly independent over $\E$ if and only if the Moore matrix $\WW_n(\aa,\phi)$ is invertible.
\end{proposition}

As a corollary we get the following result. 
\begin{corollary}\label{cor:1}
Let $\F/\E$ be a field extension and $\phi\in Aut_\E(\F)$ with $\E$ being its fixed field. If $L_\phi \in \Lf$ be a non-zero $\E$-linear operator of degree $k$, then the nullity of $L_\phi$ is at most $k$.
\end{corollary}
\begin{proof}
 If kernel of $L_\phi = \sum_{i=0}^{k}h_i\phi^i$ has dimension strictly greater than $k$, then there exists a set of $\E$-linearly independent elements $\aa=\{a_0, \ldots, a_{k}\}$ such that $(h_0, \ldots, h_k)\WW_{k+1}(\aa, \phi) = \0$. But Proposition \ref{PRO15} implies that $\WW_{k+1}(\aa,\phi)$ is invertible and thus all the $h_i$'s must be zero which contradicts with our assumption that $L_\phi$ is non-zero.
\end{proof}
 
 For the construction of MRD codes we consider the following evaluation map: 
 %. Also assume that $\E$ is the fixed field of $\phi \in Aut_\E(\F)$. 
 For $\mathbf{a} =\{a_1, \ldots, a_n\} \subseteq \F$,
\begin{equation}\label{eq:evaluation}
\begin{aligned}
\Ev_{\mathbf{a}} \colon \Lf & \longrightarrow \F^n \\
L & \longmapsto (L_{\phi}(a_1),\ldots, L_{\phi}(a_n)).
\end{aligned} 
\end{equation}
\begin{theorem}\label{thm:4}
Let $\F/\E$ be an extension of degree $m$ and $k\leq n \leq m$ be positive integers. Assume that $\mathbf{a} = \{a_1,\dots,a_n\}$ is a set $\E$-linearly independent elements of $\F$. If $\E$ is the fixed field of $\phi \in Aut_\E(\F)$, then $\C := \Ev_{\mathbf{a}}(\Lf_k)$ is an $[n,k]$ MRD code over $\F/\E$.
\end{theorem}
\begin{proof} 
Since $a_i$'s are $\E$-linearly independent, Corollary \ref{cor:1} implies that $\Ev_{\mathbf{a}}$ is injective on $\Lf_k$ and thus $\C$ has dimension $k$.  
 Corollary \ref{cor:1} also implies that, for any $L \in \Lf_k$, $rank(L_{\phi}(a_1),\ldots, L_{\phi}(a_n))$, according to the Definition \ref{def:2}, is at least $n-k+1$. Thus all the codewords of $\C$ have rank weight at least $n-k+1$. Therefore the Singleton bound indicates that $\C$ has minimum rank distance $n-k+1$.
\end{proof}

%, the minimum distance of $\C$ is $d=n-k+1$. Note that this also implies that $\Ev$ restricted on $\Lf_k$ is injective so that the dimension of $\C$ is equal to $k$ (i.e the Hypothesis $H_{dim}$).then $\C$ is a linear code over $\F$, so to find the minimum distance of $\C$ it is enough to find the minimum of $\rank\; \xx$ for any $\xx\in \C\backslash \{\0\}$. For a linear operator $L\in \Lf_k$, we know that it defines a $\K$-linear map $\F\rightarrow\F$, and 

%

As mentioned in \cite{ALR18}), the condition that $\E$ is the fixed field of $\phi \in Aut_{\E}(\F)$ is equivalent to the assumption that $\E \hookrightarrow \F$ is a cyclic Galois extension (by Artin's Lemma) since $[\F \colon \E]$ is finite. Therefore the MRD codes in the above Theorem \ref{thm:4} are essentially the \emph{generalized Gabidulin codes} defined in \cite[Definition 12]{ALR18} as follows.

\begin{definition}\cite[Definition 12]{ALR18}\label{defn:::17}
Let $\F/\E$ be a cyclic Galois extension of degree $m$ and $Aut_\E(\F)=\<\phi\>$. Let $k \leq n \leq m$ be integers and $\aa = (a_1, \dots, a_n ) \in \Fqm^n$ be a vector of $\E$-linearly independent elements. Then the generalized Gabidulin code over $\F/\E$ of length $n$ and dimension $k$ is defined as
\[
\mathcal{G}_{\phi,k}(\aa) := \{ \left(f(a_1), \dots, f(a_n) \right)\colon f(X)\in \mathcal{L}_k[X;\phi]\}.
\]
\end{definition}

%Now we come back to the construction of generalized Gabidulin codes over arbitrary fields in \cite{ALR18}.

%In fact, they have proved the following result.
%\begin{theorem}\cite[Theorem 5]{ALR18}
%Let $\F/\E$ be a field extension and let $\phi\in Aut_\E(\F)$. Then the fixed field of $\phi$ is equal to the field $\E$ if and only if $\Lf$ satisfies $H_{dim}$, i.e. non-zero linear operator of degree $k$ has kernel of dimension $k$ at most.
%\end{theorem}
\begin{remark}\label{rem:A}
In \cite{ALR18}, the property that the dimension of the kernel smaller than the degree of the operator in Corollary \ref{cor:1} is called ``\emph{Hypothesis $H_{dim}$''}. It is proved in \cite[Theorem 5]{ALR18} that Hypothesis $H_{dim}$ is equivalent to the condition that $\E$ is the fixed field of $\phi \in Aut_{\E}(\F)$. Under this condition, i.e., $\E$ being the fixed field of $\phi$, all the four different rank metrics defined for arbitrary fields become equivalent (see \cite[Proposition 5]{ALR18}).
\end{remark}
%\begin{remark}
%It is clear that the condition $H_{dim}$ implies $\E$ is the fixed field of the automorphism $\phi_{q,\l}$. Indeed, since the kernel of the linear operator $\phi-1$ has dimension at most $1$ and it cannot be of zero dimension as $1$ is in the kernel.
%\end{remark}

%Now that we are prepared to give the construction of generalized version of Gabidulin codes over arbitrary fields extension as introduced in \cite[Definition 12]{ALR18}.

In this paper we give a construction of generalized Gabidulin codes over fields of rational functions of finite characteristic. For that purpose, we define and study suitable automorphisms over the field of rational functions $\Fqm(x)$ in the next section. 
%We will use this automorphisms to give various constructions of optimal codes (including the generalized Gabidulin codes) in the remaining sections.
\section{Automorphisms of rational functions}\label{sec:4}
%In this section, we define some automorphisms of the field of rational functions $\Fqm(x)$. By fixing an automorphism, we study some properties of the extension $\Fqm(x)/ \K$ where $\K$ is the fixed field under that automorphism.  Later we will consider the linear operators using this automorphisms to construct MRD codes following the method described in the previous section.

This section is independent of other sections, but the automorphisms of rational functions we define here plays a central role in all the constructions of linear rank-metric codes in the following sections. For the rest of this paper, for a finite field $\F$ we use $\F^{\times}$ to denote the multiplicative group of $\F$.

 For any $\l\in \Fqm^\times$, we define the following map on the ring $\Fqm[x]$:
\begin{align}\label{auto}
\fql\colon\quad  \Fqm[x] &\longrightarrow \Fqm[x] \\
  \sum_{i=0}^k f_i x^i  &\longmapsto  \sum_{i=0}^k f_i^q \l^i x^i.
\end{align}

\begin{proposition}\label{pro:21}
For any $\l\in \Fqm^\times$, $\fql$ defines a ring automorphism. Moreover, if the norm $N(\l)$ of $\l$ over $\Fqm/\Fq$ has order $q-1$ in the multiplicative group $\Fq^\times$, then the set of elements of $\Fqm[x]$ fixed by $\fql$ is a ring $\R$, where
\[
\R = \left\lbrace \sum_{i=0}^k c_i \l^{-i} x^{(q-1)i} \colon c_i \in \Fq \right\rbrace.
\]
\end{proposition}

\begin{proof}
First we show that $\fql$ is a ring homomorphism. It is obvious that $\fql(1) = 1$.
 Let $f(x) = \sum_{i=0}^{k_1} f_i x^i$ and $g(x) = \sum_{i=0}^{k_2} g_i x^i$. 
 
\begin{align*}
\fql\left(f(x)g(x)\right)
& = \sum_{l=0}^{k_1+k_2} \left(\sum_{\substack{i,j \\ i+j = l}} f_i^qg_j^q\right)\l^l x^l \\
& = \sum_{l=0}^{k_1 +k_2} \left(\sum_{\substack{i,j \\ i+j = l}} f_i^q \l^i g_j^q\l^j\right) x^l \\
& = \left(\sum_{i=0}^{k_1} f_i^q \l^i x^i\right)\left(\sum_{j=0}^{k_2} g_j^q \l^j x^j\right) \\
& = \fql\left(f(x)\right) \fql\left(g(x)\right).
\end{align*}
It is also straightforward to show that 
\begin{align*}
\fql\left(f(x)+ g(x)\right) = \fql\left(f(x)\right)+ \fql\left(g(x)\right).
\end{align*}
Thus $\fql$ is indeed a ring homomorphism. Now we show that $\fql$ is in fact an automorphism of $\Fqm[x]$.
The map $\fql$ is an injection since $\l$ is non-zero. To show surjectivity of $\fql$, take any element $\sum_{i=0}^k f_i x^i$ in $\Fqm[x]$. Now from the definition of $\fql$ it is easy to see that 
\[
\fql\left( \sum_{i=0}^k f_i^{1/q} \l^{-i/q} x^i \right) = \sum_{i=0}^k f_i x^i. 
\] Therefore $\fql$ is indeed an automorphism.
 
It is well known that the set $\R$ of elements in $\Fqm[x]$ fixed by $\fql$ forms a ring. 
Next we compute the ring $\R$ for those $\l \in \Fqm^{\times}$ where $N(\l)$ has order $q-1$ in the multiplicative group $\Fq^\times$.
Suppose that $\fql\left(\sum_{i=0}^k f_i x^i\right) = \sum_{i=0}^k f_i x^i$.
Then $f_i^q\l^i = f_i$ for all $i$, and therefore, $f_i^{q-1}\l^i = 1$ whenever $f_i\neq 0$. Raising both sides of the equation $f_i^{q-1}\l^i = 1$ to the power of $\frac{q^m-1}{q-1}$, we get $N(\l)^i = 1$. Thus $i = (q-1)s_i$, for some $s_i$, as $N(\l)$ has order $q-1$. Therefore we can assume that  $f(x) = \sum_{j=0}^l f_j x^{(q-1)j}$. So $f_j^q \l^{(q-1)j} = f_j$ and this implies $f_j\l^j\in \Fq$. Thus $f(x) = \sum_{j=0}^l (f_j\l^{j})\l^{-j} x^{(q-1)j} = \sum_{j=0}^l c_j\l^{-j} x^{(q-1)j}$ for some $ c_j\in \Fq$. Conversely, it is easy to check that the polynomials of the form $\sum_{i=0}^k c_i \l^{-i} x^{(q-1)i} \text{ with } c_i \in \Fq$  are fixed by the automorphism $\fql$.
\end{proof}

We call $\R$ to be the fixed ring of the automorphism $\fql$.
Notice that the computation of the fixed ring requires the property that $N(\l)$ has order $q-1$ in $\Fq^{\times}$. 

Next we give a characterization of all such $\l \in \Fqm^{\times}$ such that its norm $N(\l)$ has order $q-1$ based on their orders. 

\begin{lemma}\label{lem::1}
For an element $\l \in \Fqm^\times$, its norm $N(\l)$ over $\Fqm/\Fq$ has order $q-1$ in the multiplicative group $\Fq^\times$ if and only if the order of $\l$ in $\Fqm^\times$ is $(q-1)s$ for some $s \neq 1$ such that $s \mid q^{m-1} + q^{m-2}+\cdots+ q+1$ and $\gcd(q-1, \frac{q^{m-1}+\cdots+1}{s})=1$.
\end{lemma}

\begin{proof}
From the definition of norm, we have the following relation 
\begin{equation}\label{EQ3}
N(\l)^{q-1} = \l^{(q^{m-1}+q^{m-2}+\ldots+1)(q-1)}=\l^{q^m-1}.
\end{equation}

First we assume that $N(\l)$ has order $q-1$ and prove the necessary part. Let the order of $\l \text{ be } r$. If $r \mid q^{m-1}+q^{m-2}+\cdots+1$, then Equation \eqref{EQ3} implies $N(\l)=1$ which contradicts with $N(\l)$ having order $q-1$. So $r = r_1 r_2$ such that $r_1 = \gcd(r, q^{m-1}+q^{m-2}+\cdots+1)$. Then it is clear that $r_2 \mid q-1$. Suppose $q^{m-1}+q^{m-2}+\cdots+1 =r_1t$ for some integer $t \neq 1$. Then $N(\l)^{r_2} = \l^{r_1tr_2} = \l^{rt} = 1$. Hence $r_2 = q-1$ and $r = (q-1)r_1$.
Set $s =r_1$. So $s | q^{m-1}+q^{m-2}+\cdots+1$. What is left to show in this part is that $\gcd(t,q-1)=1$ where $t = \frac{q^{m-1}+q^{m-2}+\cdots+1}{s}$. Suppose $\gcd(t,q-1)=d$ and $d \neq 1$. So $d \mid r$ as $r_2 = q-1$. And this implies $r_1d \mid \gcd(q^{m-1}+q^{m-2}+\cdots+1,r)$ leading to the contradiction of our assumption that $\gcd(q^{m-1}+q^{m-2}+\cdots+1,r)=r_1$. So $\gcd (t, q-1) = 1$ which proves the necessary part.

Conversely, suppose order of $\l = (q-1)s$, $s$ satisfying the given properties. If $r$ is order of $N(\l)$, then $(q-1)s \mid (q^{m-1}+q^{m-2}+\ldots+1)r$ following Equation \eqref{EQ3}. This further implies that $(q-1) \mid \frac{q^{m-1}+q^{m-2}+\ldots+1}{s} r$. From the conditions on $s$, this implies $q-1 \mid r$. Hence we get the desired result that $r =q-1$.
\end{proof}

\begin{remark}
From the above lemma it is clear that the primitive elements in $\Fqm^\times$ indeed have the property of their norm having order $q-1$ in $\Fq^\times$. But certainly there can be more elements other than the primitive elements. As the multiplicative group of a finite field is cyclic, for any divisor $s$ of $q^{m-1}+q^{m-2}+\ldots+1$, there is an element in $\Fqm^\times$ of order $(q-1)s$. Further, if we choose $s =\gcd(q-1, q^{m-1}+\ldots+1)$, then $s$ satisfies also the second condition in the statement of the above lemma.
For $q=2$, the result is trivially true in a sense norm of any element in $\F_{2^m}^\times$ has order $1$. 
\end{remark}

 The following theorem shows how $\fql$ can be extended to the field of fractions $\Fqm(x)$ of $\Fqm[x]$ for any $\l$. Furthermore, the field of rational functions has polynomial basis over the fixed field under a given automorphism $\fql$. 
 
 %We will use this automorphisms to construct rank metric codes. 

\begin{theorem}\label{thm:24}	
The automorphism $\fql$ extends naturally to the field $\Fqm(x)$ by $\fql\left(\frac{f(x)}{g(x)}\right) = \frac{\fql(f(x))}{\fql(g(x))}$. The set of elements of $\Fqm(x)$ fixed by $\fql$ forms a field $\K$, where $\K = \left\lbrace \frac{f(x)}{g(x)}\colon f(x) \in \R,\; g(x)\in \R\backslash\{0\} \right\rbrace$. Moreover, $\Fqm(x)/\K$ is a finite extension of degree  $m(q-1)$ with basis 
\[
\B = \left\lbrace a_ix^j : 1\leq i\leq m, 0\leq j\leq q-2\right\rbrace,
\]
where  $\lbrace a_1,\ldots, a_m\rbrace$ is a basis of the extension $\Fqm/\Fq$.
\end{theorem}

\begin{proof}
We only give a proof of the last part of the theorem. The remaining parts follow immediately from the definitions.

First we prove that the degree of $\Fqm(x)/\K$ is $m(q-1)$. We see that $\K$ is the fraction field of $\R$ and thus $\K = \Fq\left(\frac{x^{q-1}}{\l}\right)$. We have the inclusion of fields $\Fqm(x)\supset \Fqm\left(\frac{x^{q-1}}{\l}\right)\supset \Fq\left(\frac{x^{q-1}}{\l}\right)$. 
Therefore $\left[\Fqm(x) :\Fq\left(\frac{x^{q-1}}{\l}\right)\right] = \left[\Fqm(x): \Fqm\left(\frac{x^{q-1}}{\l}\right)\right]\left[\Fqm\left(\frac{x^{q-1}}{\l}\right): \Fq\left(\frac{x^{q-1}}{\l}\right)\right]$. The first degree on the right hand side of the equation is $q-1$, since $T^{q-1} - x^{q-1}$ is the minimal polynomial of $x$ in $\Fqm\left(\frac{x^{q-1}}{\l}\right)[T]$. It is easy to see that the second degree on the right hand side is $m$. So $\left[\Fqm(x) : \K\right] = m(q-1)$.

Finally, let us show that 
\[
\B = \left\lbrace a_ix^j \colon 1\leq i\leq m,\; 0\leq j\leq q-2\right\rbrace,
\]
 is linearly independent over $\K$. For that, it is sufficient to prove that $\B$ is linearly independent over $\R$.
 
 Suppose that 
 \[
 \sum_{j=0}^{q-2} \sum_{i=1}^m f_{i,j}(x) a_ix^j = 0,
 \]
 where $f_{i,j}(x) = \sum_{l=0}^{k} c_l^{i,j}\l^{-l}x^{(q-1)l}$ for some $c_l^{i,j}\in \Fq$.
 
 Thus,
 \[
 \sum_{j=0}^{q-2} \sum_{i=1}^m \left(\sum_{l=0}^{k} c_l^{i,j}\l^{-l}x^{(q-1)l}\right) a_ix^j = 0.
 \]
 We get
 \[
 \sum_{j=0}^{q-2} \sum_{i=1}^m \sum_{l=0}^{k} c_l^{i,j}a_i\l^{-l}x^{(q-1)l+j} = 0.
 \]
 So,
 
 \[
 \sum_{j=0}^{q-2} \sum_{l=0}^{k} \left(\sum_{i=1}^m c_l^{i,j}a_i\l^{-l}\right)x^{(q-1)l+j} = 0.
 \]
 Since, $j\leq q-2$, then all the $x^{(q-1)l+j}$ are different when $0\leq j\leq q-2$ and $0\leq l \leq k$. Therefore, each coefficients of the polynomials are equal to zero i.e.,
 \[
 \sum_{i=1}^m c_l^{i,j}a_i\l^{-l} = 0, \quad 0\leq j\leq q-2, \quad 0\leq l \leq k.
 \]
 Thus we have that $\sum_{i=1}^m c_l^{i,j}a_i = 0$, $c_l^{i,j}\in \Fq$, but we know that the $a_i$'s are linearly independent over $\Fq$, therefore the $c_l^{i,j}$'s are all equal to zero so that $f_{i,j}(x) = 0$. 
 
 Therefore $\B$ is a basis of the extension $\Fqm(x) / \K$, since $|\B| = m(q-1) = \left[ \Fqm(x) : \K \right]$.
\end{proof}

We end this section by a remark which explains the purpose of considering the particular form of automorphisms $\fql$ of $\Fqm(x)$.
\begin{remark}
The extension $\Fqm(x) / \K$ in the above theorem is a cyclic Galois extension. To see this, consider the cyclic group generated by $\phi$ and set $H = \<\phi \>$. The order of $H$ is finite as $\phi^{m(q-1)}(x) = (N(\l))^{q-1}x = x$ and also $|H| \leq [\Fqm(x) : \K]$. As the fixed field under the finite automorphism group $H$ of $\Fqm(x)$ is $\K$, the algebraic field extension $\Fqm(x) / \K$ is in fact a Galois extension. We also have $[\Fqm(x) : \K] \leq |H|$ which follows from Artin's lemma which states that for a finite group $G$ of automorphisms of a field $E$, if the fixed field of $E$ under the automorphism group $G$ is $F$, then $[E:F] \leq \mid G \mid$. Therefore, $[\Fqm(x) : \K] = |H|$ which implies that the field extension $\K \hookrightarrow \Fqm(x)$ is actually cyclic Galois extension with $\phi$ being the generator of the Galois group. As we discussed in Section \ref{sec:3}, this is the framework for constructing \emph{ generalized Gabidulin codes} over algebraic extensions of arbitrary fields of \cite{ALR18}.
\end{remark}

 In the next section we construct the generalized Gabidulin codes over the cyclic Galois extension $\Fqm(x) / \K$ using the automorphisms $\fql$.
 
\section{Constructions of maximal rank distance codes}\label{sec:5}

We start this section by reviewing the construction of generalized Gabidulin codes of \cite{ALR18} as discussed in Section \ref{sec:3} for the particular case of our interest, i.e., the rational function fields of positive characteristic. 
%{\color{red}For that we use the construction in Definition \ref{defn:::17}}. Next we reduce the codes to finite fields and give a sufficient condition for these reduced codes to be MRD codes over finite fields extension. We prove that under a set of conditions on the parameters, these reduced codes are not equivalent to twisted Gabidulin codes. Thus we get a class of codes which are non-Gabidulin MRD codes.

First we recall our choice of parameters. For the finite extension $\Fqm/\Fq$ of degree $m$, let $\l\in \Fqm$ such that $N(\l)$ of has order $q-1$ in the multiplicative group $\Fq^\times$. We fix one such $\l$ and let $\phi = \fql$, where $\fql$ is the automorphism of $\Fqm(x)$ as given in Theorem \ref{thm:24}, i.e., $\f(\sum_{i=0}^{t} f_i x^i) = \sum_{i=0}^{t} f_i^q \l^ix^i$ for any $\sum_{i=0}^{t} f_i x^i\in \Fqm[x]$. 

In Theorem \ref{thm:24}, we also have seen that the fixed field $\K$ of $\phi$ is equal to $\Fq\left(\frac{x^{q-1}}{\l}\right)$ and $\Fqm(x)/\K$ is a cyclic Galois extension of degree $m(q-1)$.
%\[
%\K = \left\lbrace \frac{f(x)}{g(x)}\colon f(x) \in \R,\; g(x)\in \R\backslash\{0\} \right\rbrace,
%\]
%where
%\[
%\R = \left\lbrace \sum_{i=0}^k c_i \l^{-i} x^{(q-1)i} \colon c_i \in \Fq \right\rbrace.
%\]
Here we define the generalized Gabidulin codes over rational function fields.
\begin{definition}\label{def:ourcode}
Let $q$ be a prime power and $k,m,n$ be positive integers such that $n=m(q-1)$ and $k \leq \min\{m,n\}$. Let $\l \in \Fqm^{\times}$ such that $N(\l)$ has order $q-1$ in the multiplicative group $\Fq^{\times}$. Also assume $\phi = \fql$, where $\fql$ is the automorphism of $\Fqm(x)$ defined as $\f(\sum_{i=0}^{t} f_i x^i) = \sum_{i=0}^{t} f_i^q \l^ix^i$ and let $\K$ be the fixed field of $\phi$. Then the generalized Gabidulin code over $\Fqm(x)/\K$ of length $n$ and dimension $k$ is defined as 
    \[\C_{\l,k}(\B):= \{(L_{\phi}(\alpha_1),L_{\phi}(\alpha_2)\dots L_{\phi}(\alpha_{m(q-1)}) )\colon L_{\phi} \in \Lf_k\},\]
    where $\B = \left\lbrace \alpha_1, \ldots, \alpha_{m(q-1)}\right\rbrace$ is a basis of $\Fqm(x)$ over $\K$.
\end{definition}

\begin{remark}
It is to be noted that in the above definition one can take the length $n$ of $\C_{\l,k}(\B)$ to be less than $ m(q-1)$ also. But we consider the maximum length $m(q-1)$ throughout for all the constructions of optimal rank-metric codes using the MRD codes $\C_{\l,k}(\B)$.
\end{remark}

In Theorem \ref{thm:24}, we have seen that $\Fqm(x)$ admits a polynomial basis over $\K$ given by
\[
\B = \left\lbrace a_ix^j : 1\leq i\leq m, 0\leq j\leq q-2\right\rbrace,
\]
 where $\{a_1, \ldots, a_m\}$ is an $\Fq$-basis of $\Fqm$.
We will observe that in all the constructions in this paper this polynomial basis plays a central role. We record below the form of a generator matrix for the code $\C_{\l,k}(\B)$ which will be useful in all our constructions.

\begin{proposition} \label{PRO25}
%If $\Lf_k\subset \Lf$  is the set of all linear operators of degree $k-1$ at most, then  $\C = \Ev(\Lf_k)$ is an $[n,k,d]$-linear MRD code over $\Fqm(x)/\K$. The generator matrix of $\C$ has the form,
The code $\C_{\l,k}(\B)$ of Definition \ref{def:ourcode} is an $[n,k]$-MRD code over $\Fqm(x)/\Fq\left(\frac{x^{q-1}}{\l}\right)$ with a generator matrix $G$ of the following form
{\footnotesize
\begin{equation}\label{eq:3}
 \begin{pmatrix}
g_{1,1,0} & \ldots & g_{1,m,0} & g_{1,1,1} x & \ldots & g_{1,m,1} x & \ldots & g_{1,1,q-2} x^{q-2} &\ldots & g_{1,m,q-2} x^{q-2} \\
g_{2,1,0} & \ldots & g_{2,m,0} & g_{2,1,1} x & \ldots & g_{2,m,1} x & \ldots & g_{2,1,q-2} x^{q-2} &\ldots & g_{2,m,q-2} x^{q-2} \\
\vdots    & \ddots & \vdots    & \vdots      & \ddots & \vdots      & \ddots & \vdots              &\ddots & \vdots              \\
g_{k,1,0} & \cdots & g_{k,m,0} & g_{k,1,1} x & \ldots & g_{k,m,1} x & \ldots & g_{k,1,q-2} x^{q-2} &\ldots & g_{k,m,q-2} x^{q-2} \\
\end{pmatrix},
\end{equation}
}
where $g_{t,u,v}\in \Fqm$ for $1\leq t\leq k$, $1\leq u\leq m$ and $0\leq v \leq q-2$.
\end{proposition}
\begin{proof}
It is easy to see that $\C_{\l,k}(\B) = \Ev(\Lf_k)$ where $\Ev$ is the evaluation map $\Ev: \Lf \rightarrow \Fqm(x)^n$ defined by $\B$ as given in equation \eqref{eq:evaluation}.
%\[
%L  \mapsto (L(a_1),\dots,L(a_m)|L(a_1 x),\dots,L(a_m x)|\dots|L(a_1 x^{q-2}),\dots,L(a_m x^{q-2})).
%\]
It directly follows from Theorem \ref{thm:4} that $\C_{\l,k}(\B)$ is an MRD code since the fixed field $\K$ of $\phi=\fql$ is $\Fq\left(\frac{x^{q-1}}{\l}\right)$ as implied by Theorem \ref{thm:24}.
\end{proof}

%We will use the Gabidulin code over the rational functions in Proposition \ref{PRO25} to construct new class of codes in different setting. In the remaining part of this section, we show that for some $q, m, k$ with $k\leq m$, we can construct MRD codes which are not equivalent to twisted Gabidulin codes by reducing the code over rational functions over a finite field.

%{\color{green} Do we need to write this paragraph here?} Notice that the usual rank weight we consider on the code $\C_{\l,k}(\B)$ is given by the dimension over $\K$ of the $\K$-subspace of $\Fqm(x)$ generated by the entries of the codeword. 
%We will also be interested in the dimension of the $\Fq$-subspace of $\Fqm(x)$ generated by these entries. The reason for this is that, we will work with the codewords of $\C_{\l,k}(\B)$ with entries in the polynomial ring. These codewords over polynomial ring can for example be reduced to a codewords over some extension of $\Fqm$ or we can also use them to construct matrix rank-metric codes with specific properties.
\begin{remark}\label{REM5}
Since the automorphism $\fql$ of $\Fqm(x)/\K$ has fixed field $\K$, then we can use the decoding algorithm of generalized Gabidulin codes over arbitrary field in \cite{ALR18} with our code. The algorithm has complexity $\mathcal{O}(N^2)$ operations over $\Fqm(x)$, if $N$ is the length of the code. In general it is not easy to have an exact complexity of the decoding algorithm when considering operations over $\Fqm$. Indeed, this depends on the degree of the polynomials in the received word. The decoding algorithm involves some multiplications of polynomials and each multiplication increases the degree of the polynomials being involve while decoding.
\end{remark}

\subsection{Reduction to finite fields}
In this section, we construct MRD codes over finite fields by reducing the generalized Gabidulin codes over rational function fields $\Fqm(x)/\K$ modulo suitable prime ideal in $\Fqm[x]$. 
%Note that if $\GG$ is a generator matrix of a rank-metric code over the extension $\Fqm(x)/\K$ given in Proposition \ref{PRO25}, then $\det \GG\MM\neq 0$ for any $\MM \in \K^{(n\times k)}$ of rank $k$. In particular, for any $\MM \in \Fq^{(n\times k)}$ of rank $k$, $\det \GG\MM\neq 0$. In the following theorem we show that this is a necessary condition for the reduced code to be MRD.
%If, we are to get an MRD code after reducing to finite fields, we must keep this determinant non-zero.

We will work only on codes of length $n=m(q-1)$ over the extension $\Fqm(x)/\K$ with generator matrix as given in Proposition \ref{PRO25}. A construction of codes of smaller length can be done by choosing fewer columns of the generator matrix.

%By Theorem \ref{thm:3}, since $\GG$ is a generator matrix of an MRD code over $\F/\K$, then for any $(n\times k)$-matrix $\MM$ over $\K$ of rank $k$, $\det \GG\MM\neq 0$. In particular, for any $(n\times k)$-matrix $\MM$ over $\Fq$ of rank $k$, $\det \GG\MM\neq 0$. Now, to ensure that Theorem \ref{thm:26} gives us an MRD codes, we must choose $f(x)$, so that the reduction are non-zero.
\begin{theorem}\label{thm:26}
Suppose that $\GG$ is the matrix in $\Fqm[x]^{k\times n}$ given in Proposition \ref{PRO25} and consider $\G$ to be the set of all $(n\times k)$-matrices over $\Fq$ in column reduced echelon form. Let $f(x)\in \Fqm[x]$ be an irreducible polynomial of degree $r \geq q-1$. We denote by $\overline{\GG}$ the matrix where its entries are those of $\GG$ modulo $f(x)$. Then $\overline{\GG}$ generates an MRD code over $\F_{q^{mr}}/\Fq$ if and only if $\det \GG\MM\not\equiv 0 \mod (f(x))$ for any $\MM\in \G$.
\end{theorem}
\begin{proof}
By taking the entries of $\GG$ modulo $f$, we see that $\overline{\GG}$ defines a linear rank-metric code of length $m(q-1)$ over $\F_{q^{mr}}\simeq\Fqm[x]\mod f(x)$. By Theorem \ref{thm:3}, $\overline{\GG}$ defines an MRD code if and only if $\det\overline{\GG}\MM\neq 0$ for any $\MM \in \Fq^{n\times k}$ of rank $k$. We can restrict our choices of $\MM$ to matrices in column reduced echelon form as any $\MM$ can be transformed into a matrix in column reduced echelon form without changing the rank of $\overline{\GG}\MM$.
\end{proof}

\begin{remark}
\begin{enumerate}
    \item In Theorem \ref{thm:26}, we would like to have $r$ to be the smallest possible (and preferably $r=q-1$) so that the length of the code is equal to the degree of the extension $\F_{q^{mr}}/\Fq$. 
In general it is not clear what is the minimum possible $r$. 
However, $r=k(q-2)+1$ would guarantee that the polynomial $f(x)$ does not divide any of the determinants $\det \GG\MM$. But in that case
the code will have length $m(q-1)$, while the base field is $\F_{q^{m(k(q-2)+1)}}$. 
The length of the code is roughly a factor $k$ smaller than the extension degree of the code, which is significantly smaller compared to other explicit MRD constructions. 

\item Of course, one may consider $r$ to be smaller, but there is no theoretical proof that the reduced code will be MRD. 
The only strategy to check if the code is MRD is to brute force the computation of all the determinant $\det\overline{\GG}\MM$. 
This is not practical as the number of determinant to be computed is very large, roughly, $\begin{bmatrix}
n \\
k
\end{bmatrix}_q$.

\end{enumerate}
\end{remark}
However, this construction provides evidence for the existence of many rank-metric codes exist when the base field is large enough. In fact, we show later that the reduced codes have a very interesting property. First, we want to give an example with small parameters.

\begin{example}\label{EXA20}
Let $q=m=k=3$. A basis of the extension $\F_{3^3}[x]/\K$ is given by $\lbrace 1,a,a^2,x,ax,a^2x\rbrace$, where $a\in \F^{3^3}$ such that $a^3-a+1=0$. Suppose that $\l=-1$ such that the automorphism $\phi$ is defined by $\phi(\a x^i)= (-1)^i\a^q x^i$ for $\a\in \Fqm$ and $i$ a non-negative integer. The generator matrix of the code over $\F_{3^3}[x]/\K$ is 
\[
\GG = \begin{pmatrix}
1 & a & a^{2} & x & a x & a^{2} x \\
1 & a + 2 & a^{2} + a + 1 & 2 x & \left(2 a + 1\right) x
& \left(2 a^{2} + 2 a + 2\right) x \\
1 & a + 1 & a^{2} + 2 a + 1 & x & \left(a + 1\right) x
& \left(a^{2} + 2 a + 1\right) x
\end{pmatrix}
\]
\end{example}
Suppose that $\F_{3^{12}}=\F_{3^3}(b)$ where 
\[
b^{4} + \left(2 a^{2} + a + 1\right) b^{3} + \left(a^{2} + 2\right)
b^{2} + \left(2 a^{2} + a + 1\right) b + a + 1 = 0.
\]
By Theorem \ref{thm:26}, the generator matrix 
\[
\overline{\GG} = \begin{pmatrix}
1 & a & a^{2} & b & a b & a^{2} b \\
1 & a + 2 & a^{2} + a + 1 & 2 b & \left(2 a + 1\right) b
& \left(2 a^{2} + 2 a + 2\right) b \\
1 & a + 1 & a^{2} + 2 a + 1 & b & \left(a + 1\right) b
& \left(a^{2} + 2 a + 1\right) b
\end{pmatrix}
\]
generates an MRD code of length $n=6$ and dimension $k=3$ over the extension $\F_{3^{12}}/\F_3$.

If $\C$ is the code generated by $\overline\GG$, it can be shown that any $q^s$-th power of $\C$ intersects with the code $\C$ only trivially. By Proposition \ref{pro:3} we can conclude that the code is not a generalized twisted Gabidulin code and thus not equivalent to any of these Gabidulin type codes.

The above example poses a natural question whether the reduction over finite fields of the newly constructed MRD codes over rational functions can provide a large class of MRD codes which are not equivalent to generalized twisted Gabidulin codes. Indeed, as the next theorem shows, the observation in the above example is not something sudden, but this is true for a more general set of parameters $q$, $m$, $k$ with $k\leq m$ and large enough $r$.

\begin{theorem}\label{thm:28}
Let $q\geq 3$ be a prime power and $m,n,k$ be positive integers such that $k \leq m$, $n=(q-1)m$, and $q-1 | m$. Let $\C := \Ev(\mathcal{L}(\fql))$ be the $[n,k]$ MRD code over $\Fqm(x)$ where $\l \in \Fqm^\times$ and that $N_{\Fqm/\Fq}(\l)$ generates the multiplicative group $\Fq^\times$. For $r = k(q-2)+1$, let $\tilde{\C}$ be the MRD code obtained by reduction of $\C$ to the finite field $\F_{q^{mr}}$. Now for any integer $s$ with $\gcd(s,mr) = 1$, we have $\tilde{\C} \cap \tilde{\C}^{q^s} = \{0\}$, where
\[
\tilde{\C}^{q^s}= \{(c_1^{q^s}, \ldots,c_n^{q^s}): (c_1,\ldots,c_n) \in \C \}.
\]
\end{theorem}

\begin{proof}
First we prove the result for $q=3$ and $k=m$.

Let $\{a_1,\dots,a_m\}$ be an $\Fq$-basis of ${\Fqm}$ and we set $\l^{*(i)} = \Pi_{j=0}^{i-1}{\l^{q^j}} = \l^{\sum_{j=0}^{i-1} q^j}$ and $\l^{*0} = 1$. We consider the following generator matrix of the code $\C$: 
\[
\AA = \begin{pmatrix}
a_1 & \dots & a_m & a_1 x & \dots & a_m x \\
a_1^q & \dots & a_m^q & a_1^q \l^{*(1)} x & \dots & a_m^q \l^{*(1)} x \\
\vdots & \ddots & \vdots & \vdots & \ddots & \vdots \\
a_1^{q^{m-2}} & \dots & a_m^{q^{m-2}} & a_1^{q^{m-2}} \l^{*(m-2)} x & \dots & a_m^{q^{m-2}} \l^{*(m-2)} x \\
a_1^{q^{m-1}} & \dots & a_m^{q^{m-1}} & a_1^{q^{m-1}} \l^{*(m-1)} x & \dots & a_m^{q^{m-1}} \l^{*(m-1)} x \\
\end{pmatrix}.
\]

We choose an irreducible polynomial $h(x)$ in $\Fqm[x]$ of degree $r=k(q-2)+1$. Suppose $h(b)=0$ where $b\in \Fqmr\simeq \Fqm(x)/(h(x)$. By Theorem \ref{thm:26} we get an MRD code $\tilde{\C}$ over $\Fqmr$ with generator matrix
\[
\GG = \begin{pmatrix}
a_1 & \dots & a_m & a_1 b & \dots & a_m b \\
a_1^q & \dots & a_m^q & a_1^q \l^{*(1)} b & \dots & a_m^q \l^{*(1)} b \\
\vdots & \ddots & \vdots & \vdots & \ddots & \vdots \\
a_1^{q^{m-2}} & \dots & a_m^{q^{m-2}} & a_1^{q^{m-2}} \l^{*(m-2)} b & \dots & a_m^{q^{m-2}} \l^{*(m-2)} b \\
a_1^{q^{m-1}} & \dots & a_m^{q^{m-1}} & a_1^{q^{m-1}} \l^{*(m-1)} b & \dots & a_m^{q^{m-1}} \l^{*(m-1)} b \\
\end{pmatrix}.
\]

If we write $G = (M|N)$ where both matrices $M,N$ have $m$ columns. Then $N = bD\MM$, where 
\[ D = \text{diag}(\l^{*(i)} ~| ~ i= 0, \ldots, m-1).\]

%\[
%D = \begin{pmatrix}
%&1      &\quad &\quad &\quad \\
%&\quad  &\l^{*1} &\quad   \\
%&\quad &\quad &\ddots &\quad\\
%&\quad &\quad &\quad  &\l^{*(m-1)}
%\end{pmatrix}.
%\]
Therefore, $\GG^{[s]}= (M^{[s]} | N^{[s]} )$, where $\GG^{[s]}$ is the matrix obtained from $\GG$ by raising all entries to the power $q^s$. Note that $M$ is a Moore matrix of order $m$ and raising the entries of $\MM$ by a power of $q$ only shifts the rows upwards (and cyclically) $s$ times. Thus we can write $\MM^{[s]} = \PP^s\MM$, where $\PP$ is permutation matrix,
%\[
%\PP = \begin{pmatrix}
%0 & 1 & 0 & \hdots & 0 \\
%0 & 0 & 1 & \ddots & 0 \\
%0 & \ddots & \ddots & \ddots & 0 \\
%0 & \ddots & \ddots & \ddots & 1 \\
%1 & 0 & \hdots & \hdots & 0 \\
%\end{pmatrix}.
%\]

\[\PP = \begin{pmatrix}
           0 &I_{m-1}\\
           1 &0
          \end{pmatrix} 
          \]
We want to show that $\dim_{\Fqmr} \tilde{\C}\cap \tilde{\C}^{q^s} = 0$ for any $s$ where $\gcd(mr,s)=1$.

Assume that the intersection is nonempty. So there are $\ff=(f_1,\dots,f_{m})$ and $\g=(g_1,\dots,g_{m}) \in \Fqmr^k$ such that $\ff\GG = \g \GG^{[s]}$. Thus $\ff M = \g \PP^{[s]}M $ and $\ff = \g  \PP^{[s]}$ by injectivity of $M$.

\begin{equation}\label{1}
\ff\GG = \g \GG^{[s]}  \Rightarrow  \ff M = \g \PP^{[s]}M.
\end{equation}
%where $\GG^{[s]}$ is the matrix obtained from $\GG$ by raising all entries to the power $q^s$. Notice that the first $m$ columns of the matrix $\GG$ is a Moore matrix of order m and we name it $\MM$. Now considering the first $m$ columns of Equation \eqref{1}, we get 
%
%\begin{equation}\label{2}
%\ff\MM = \g\MM ^{[s]},
%\end{equation}
%where 
%
%As raising the entries of $\MM$ by a power of $q$ only shifts the rows upwards (and cyclically) $s$ times, we can write $\MM^{[s]} = \PP^s\MM$, where $\PP$ is permutation matrix,
%\[
%\PP = \begin{pmatrix}
%0 & 1 & 0 & \hdots & 0 \\
%0 & 0 & 1 & \ddots & 0 \\
%0 & \ddots & \ddots & \ddots & 0 \\
%0 & \ddots & \ddots & \ddots & 1 \\
%1 & 0 & \hdots & \hdots & 0 \\
%\end{pmatrix}.
%\]
%So Equation \eqref{2} becomes

%\begin{equation}\label{EQ7}
%\ff\MM = \g\PP^s\MM \Rightarrow \ff=\g\PP^s \text{ as }\MM \text{ is invertible}.
%\end{equation} 
%Next we consider the submatrix of $\GG$ consisting of the last $m$ columns and denote it by $N$.

Furthermore, $\ff b D M = \g b^{[s]}  D^{[s]} M^{[s]}$, thus $ \g b\PP^s D = \g b^{[s]} D^{[s]}\PP^s$ or $\g (b \PP^s D - b^{q^s} D^{[s]}\PP^s)= 0$.

%Therefore, Equation \eqref{1} gives
%\begin{align*}
%\quad b \ff D\MM &= b^{q^s} \g D^{[s]}\MM^{[s]} \\
%\Rightarrow b\ff D\MM &= b^{q^s}\g D^{[s]}\PP^s\MM \\
%\Rightarrow \quad b\ff D &= b^{q^s}\g D^{[s]}\PP^s \\
%\Rightarrow b\g\PP^s D &= b^{q^s}\g D^{[s]}\PP^s \text{ (using Equation }\eqref{EQ7})
%\end{align*}

 With $s = m z + t$ for some nonnegative integers $z$ and $t$ with $0 \leq t<m$, we have $\PP^s = \PP^t$, and furthermore, 
 \[\PP^t= \begin{pmatrix}
          0 &I_{m-t}\\
          I_t &0
              \end{pmatrix}.             
 \]

This gives us 
\[b\PP^sD - b^{q^s}D^{[s]}\PP^s = \begin{pmatrix}
                                   0 &A \\
                                   B &0
                                   \end{pmatrix},
\]
where
\begin{align*}
 A &= \text{diag}(b\l^{*(t+i-1)} - b^{[s]}(\l^{[s]})^{*(i-1)} ~|~ i = 1, \ldots, m-t), \\
 B &= \text{diag}(b\l^{*(i-1)}  - b^{[s]}(\l^{[s]})^{*(m-t+i-1)} ~|~ i = 1, \ldots, t).
 \end{align*}
%$\PP^t = (P_{i,j})_{1\leq i,j\leq m}$ is therefore described by
%\[
%\begin{cases}
%P_{m-t+j,j} = 1, & \text{ for } j=1,\dots,t,\\
%P_{j-t,{j}} = 1, & \text{ for } j=t+1,\dots,m,\\
%P_{i,j} = 0, &\text{ otherwise}.
%\end{cases}
%\]

%If $b\g\PP^sD =(u_1,\dots,u_m)$, then
%\[
%\begin{cases}
%u_j = b g_{m-t+j} \l^{*(j-1)}, & \text{ for } 1\leq j \leq t\\
%u_j = b g_{j-t}\l^{*(j-1)}, & \text{ for } t+1\leq j\leq m.
%\end{cases}
%\]
%On the other side, if $b^{q^s}\g D^{[s]}\PP^s = (v_1,\dots,v_m)$. 
%\[
%\begin{cases}
%v_j = b^{q^s} g_{m-t+j} \l^{*(m-t+j-1)}, & \text{ for } 1\leq j \leq t\\
%v_j = b^{q^s} g_{j-t}\l^{*(j-t-1)}, & \text{ for } t+1\leq j\leq m.
%\end{cases}
%\]
%
%Thus $b\g\PP^s D = b^{q^s}\g D^{[s]}\PP^s$ implies
%\[
%\begin{cases}
%b g_{m-t+j} \l^{*(j-1)}= b^{q^s} g_{m-t+j} (\l^{*(m-t+j-1)})^{q^s}, & \text{ for } 1\leq j \leq t\\
%b g_{j-t}\l^{*(j-1)}= b^{q^s} g_{j-t}(\l^{*(j-t-1)})^{q^s}, & \text{ for } t+1\leq j\leq m.
%\end{cases}
%\]
Now we show that the matrices $A$ and $B$ are nonsingular. First, let $b \l^{*(i-1)}= b^{q^s} \l^{*(m-t+i-1)}$ for some $i$ such that $1\leq i\leq t$. By taking the norm of the elements on the both side of the equation, we get $N_{\Fqm/\Fq}(\l)^{i-1} = N_{\Fqm/\Fq}(\l)^{m-t+i-1}$ i.e. $N_{\Fqm/\Fq}(\l)^{m-t}=1$. Since $N(\l)$ has order $q-1$, then $q-1$ divides $m-t$ and this combining with our assumption that $q-1 | m$ implies that $(q-1)$ divides $\gcd(mr,s)$. But we assumed $\gcd(mr,s) = 1$ and therefore, $B$ is nonsingular.

Similarly, if $b \l^{*(t+i-1)} = b^{[s]} \l^{*(i-1)}$ for $1\leq i\leq m-t$, then $N(\l)^{t+i-1} = N(\l)^{i-1}$. So $N(\l)^t=1$ and thus $q-1$ divides $t$. As explained above, this leads to a contradiction with $\gcd(mr,s) =1$. So $A$ is also nonsingular. And nonsingularity of $A,B$ implies that $\g = \0$ or $\ff = \g \PP^s = \0$. Thus it proves that $\dim_{\Fqmr} \tilde{\C}\cap \tilde{\C}^{q^s} = 0$ for the case $k = m$.

 Now suppose $k <m$. We consider $\GG'$ to be the $m \times (q-1)m$ matrix which is obtained from $\GG$ by appending $m-k$ rows at the end as follows:
{\small
 \[
\GG' = \left(
\begin{array}{cccccccc}
&  & & \GG  & & & \\ \hline
a_1^{q^{m-2}} & \dots & a_m^{q^{m-2}} & a_1^{q^{m-2}} \l^{*(m-2)} b & \dots & a_m^{q^{m-2}} \l^{*(m-2)} b \\
a_1^{q^{m-1}} & \dots & a_m^{q^{m-1}} & a_1^{q^{m-1}} \l^{*(m-1)} b & \dots & a_m^{q^{m-1}} \l^{*(m-1)} b \\
\end{array}
\right).
\]
}
 
 We can rewrite $(f_1,\dots,f_{k})G = (g_1,\dots,g_{k})G^{[s]}$ as
\begin{equation}\label{1'}
(f_1,\dots,f_{k}, 0, \ldots, 0) \GG' = (g_1,\dots,g_{k}, 0, \ldots, 0) \GG'^{[s]}.
\end{equation} 
 Then rest of the proof is same as for the case $k=m$.
 
 For $q>3$, if we have $k=m$ then we prove the result by applying the same proof for $q=3$ as there we only need to consider the submatrix of the generator matrix consists of the first $2m$ columns. And the case of $k<m$ can be solved as described in the above paragraph. 
\end{proof}
 Combining the above theorem with Proposition \ref{pro:3}, we get the following corollary.
\begin{corollary}\label{cor:new}
Let $q\geq 3$ be a prime power and $m,n,k$ be positive integers such that $k \leq m$, $n=(q-1)m$, and $q-1 | m$. Let $\C := \Ev(\mathcal{L}(\fql))$ be the $[n,k]$ MRD code over $\Fqm(x)$ where $\l \in \Fqm^\times$ and that $N_{\Fqm/\Fq}(\l)$ generates the multiplicative group $\Fq^\times$. For $r = k(q-2)+1$, if $\tilde{\C}$ be the MRD code obtained by reduction of $\C$ to the finite field $\F_{q^{mr}}$, then $\tilde{\C}$ is not equivalent to any generalized twisted Gabidulin codes. 
\end{corollary}

We would like to mention that the codes $\tilde{\C}$ in Corollary \ref{cor:new} are not equivalent to many existing constructions, e.g., those in \cite{PRS17}. In \cite{PRS17}, the intersection $\C\cap \C^{q^s}$ surely can have dimension smaller than $\dim \C-2$ but the intersection is not trivial. 

 In \cite{ALR18}, the authors studied the generalized Gabidulin codes over number fields and they proved that the reduction of these codes to finite fields gives back the Gabidulin codes (over finite fields).

In fact, the large intersection of a Gabidulin code $\C$ with $\C^{q^s}$ is a key property exploited by Overbeck \cite{Ove08} to fully break the Gabidulin-Paramonov-Tretjakov (GPT) cryptosystem based on Gabidulin codes \cite{GPT91}. That is also a reason why twisted Gabidulin codes were not considered to be secure for such GPT cryptosystem. In our case, we have codes where the intersection is trivial. 
%In the next section, we give another application of the MRD codes constructed over rational function fields.

\section{Constructions of optimal Ferrers diagram rank-metric codes}\label{sec:6}
In this section we consider matrix rank-metric codes (Definition \ref{def:29}). In \cite{ES09}, Etzion and Silberstein proposed the use of certain matrix rank-metric codes called Ferrers diagram rank-metric (FDRM) codes to get large subspace codes. 
We give constructions of some optimal Ferrers diagram rank-metric codes using the generalized Gabidulin codes over rational function fields. 

Let $\Fq^{m\times n}$ denotes the $\Fq$-vector space of ${m\times n}$ matrices with entries in $\Fq$. For an $\Fq$-basis of $\Fqm$, say $\beta = \{ \alpha_1, \ldots, \alpha_m\}$, the following map gives an $\Fq$-vector space isomorphism between $\Fqm^n$ and $\Fq^{m\times n}$: 
\begin{equation}\label{eq:psi}
\psi_{\beta}: (x_1, \ldots, x_n) \mapsto ([x_1]_\beta, \ldots, [x_n]_\beta)
\end{equation}
where $[x_i]_\beta = $ 
$
\begin{bmatrix}
a_1\\
a_2\\
\vdots\\
a_m
\end{bmatrix}
 \in \Fq^{m \times 1}$ if $x_i = \sum_{i=1}^{m}a_i \alpha_i$. 
 
  Note that $d(A,B) := rank(A-B)$ for $A,B \in \Fq^{m \times n}$ defines a metric on the space $\Fq^{m \times n}$ and this further implies that $\psi_{\beta}$ is an isometry if we consider the \emph{rank metric} on $\Fqm^n$ as in Definition \ref{def:2}.
 
 \begin{definition}\label{def:29}
 Any $\Fq$-subspace $\C$ of $\Fq^{m \times n}$ with the induced metric is called a matrix rank-metric code. If $d = \min\{rank(A): A \in \C\backslash \{0\}\}$ and $\dim \C = k$, then we say that $\C$ is an $[m\times n,k,d]$-rank-metric code.
 \end{definition}

  We recall some notions related to rank-metric codes in Ferrers diagrams.

 \begin{definition}\label{defn:30}
An $m \times n$ Ferrers diagram $\CF$ is an array of dots and empty entries with the following properties:
\begin{itemize}
\item the number of dots in each row is at most the number of dots in the previous row.
\item all the dots are shifted to the right.
\item the first row has $n$ dots and there are $m$ dots in the last column.
\end{itemize}
\end{definition}

So an $m \times n$ Ferrers diagram can be visualized as a tuple $\{c_1,\ldots, c_n\}$ with $c_1 \leq c_2 \leq \ldots \leq c_n$ where $c_i$ is the number of dots in the $i$-th column such that all the dots are right and top aligned and $c_n = m$.

\begin{definition}[Rank-metric codes in Ferrers diagram]
Let $\C \subseteq \Fq^{m \times n}$ be a rank-metric code and $\CF$ be an $m \times n$ Ferrers diagram. If every codeword in $\C$ has shape $\CF$, i.e., the nonzero entries of any codeword are only in those positions with dots in the Ferrers diagram $\CF$, we call $\C$ to be a Ferrers diagram rank-metric code. We use the notation $[\CF,k,d]_q$ for the rank-metric codes supported by $\CF$ with dimension $k$ and minimum distance $d$. 
\end{definition}

\begin{example}
Let $\CF = \{ 2,2,3,5\}$ be a Ferrers diagram which can be visualized as a right and top aligned array of dots as given below.
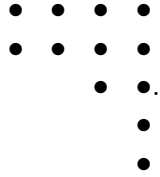
\[
\begin{matrix}
\bullet & \bullet & \bullet & \bullet\\
\bullet & \bullet & \bullet & \bullet\\
 &  & \bullet & \bullet\\
 &  &  & \bullet\\
 &  &  & \bullet\\
\end{matrix}.
\]
The matrix space over $\F_2$ given by
\[
\C = \left< 
\begin{pmatrix}
1 & 0 & 1 & 1\\
1 & 1 & 1 & 0\\
0 & 0 & 1 & 1\\
0 & 0 & 0 & 1\\
0 & 0 & 0 & 1\\
\end{pmatrix} ,
\begin{pmatrix}
1 & 0 & 1 & 1\\
0 & 0 & 1 & 0\\
0 & 0 & 1 & 1\\
0 & 0 & 0 & 0\\
0 & 0 & 0 & 1\\
\end{pmatrix}
\right>
\]
is an $[\CF,2,2]_2$-Ferrers diagram code.
\end{example}

In \cite{ES09}, Etzion and Silberstein presented an upper bound on the dimension of rank-metric codes in Ferrers diagram over arbitrary fields.

\begin{theorem}[\cite{ES09}]\label{THM27}
Let $\CF =\{r_1, r_2, \dots, r_n\}$ be a Ferrers diagram and $\C$ be an $[\CF, k, d]$-Ferrers diagram code. If we define
\begin{align*}
v_i &=\left\{
 \begin{array}{l}
 \text{number of dots remaining after removing the} \\ 
 \text{top $i$ rows and $d-1-i$ columns from right} 
 \end{array}
\right\}  \\
&=\sum_{j=1}^{n-d+1+i} \max \{0, r_j-i\},
\end{align*}
 then 
$
k \leq \min_{0 \leq i \leq d-1}v_i.
$
\end{theorem}
We use $v_{min}[\CF, d]$ to denote the upper bound $\min_{0 \leq i \leq d-1}v_i$ in the above theorem for a given Ferrers diagram $\CF$ and a target distance $d$. 

\begin{definition}
 An $[\CF,k, d]$ rank-metric code $\C$ is called optimal if $k= v_{\min}[\CF,d]$.  
\end{definition}
About the existence of optimal Ferrers diagram rank-metric codes over finite fields, Etzion and Silberstein formulated a conjecture in \cite{ES09}.
\begin{conj}[\cite{ES09}]
For every $m \times n$ Ferrers diagram $\CF$ and every $1 \leq d \leq \min\{m,n\}$, there exists optimal $[\CF, d]_q$ codes for every finite field $\Fq$. 
\end{conj}

Many constructions of optimal Ferrers diagram codes are already known \cite{ES09,EGRWZ16,GR17,AG19,LCF19,ZG19}. There are constructions which are based on maximum distance separable (MDS) codes, subcodes of MRD codes and forming new codes by combining known Ferrers diagram codes.
For a survey of known constructions of Ferrers diagram rank-metric codes, one can see \cite{LCF19}. 

In this section, we present constructions of optimal FDRM codes. The Ferrers diagrams we consider are motivated by the Ferrers diagram $\CF =\{2,2,4,4,6,6\}$ mentioned in \cite[Section VIII]{EGRWZ16} by Etzion et al. They asked whether there are optimal rank-metric codes in $\CF$ (in Figure \ref{fig::1}) with minimum distance $d=4$ and any prime power $q$: 

\begin{figure}[hbt!]
\[
\begin{matrix}
 \bullet & \bullet & \bullet & \bullet & \bullet & \bullet \\
 \bullet & \bullet & \bullet & \bullet & \bullet & \bullet \\
 & & \bullet & \bullet & \bullet & \bullet \\
 & & \bullet & \bullet & \bullet & \bullet \\
 & & & & \bullet & \bullet \\
 & & & & \bullet & \bullet \\
\end{matrix}.
\]
\caption{$\CF =\{2,2,4,4,6,6\}$}\label{fig::1}
\end{figure}

This is answered affirmatively in \cite[Example III.16]{AG19}. We are interested in constructing optimal FDRM codes for Ferrers diagrams which generalize the pattern of $\CF =\{2,2,4,4,6,6\}$. We consider generalizations in three steps and construct optimal FDRM codes in each step. Our constructions are based on the MRD codes over rational functions constructed in previous section and representation of elements in finite fields.

%Here we consider a Ferrers diagram which is generalizing the pattern given in Figure \ref{fig::1}. Our construction of optimal rank metric codes in this Ferrers diagram is 

Before describing our Ferrers diagrams, we first note that any $m \times n$ Ferrers diagram $\CF$ can be written as 
\[
\CF = \{\overbrace{r_1,\ldots,r_1}^{{\uu}_1\text{ times }},\overbrace{r_2,\ldots,r_2}^{{\uu}_2\text{ times }},\dots,\overbrace{r_p,\ldots,r_p}^{{\uu}_p\text{ times }}\},
\]
where $r_i < r_j$ for $1 \leq i < j \leq p$, $\sum_{i = 1}^p {{\uu}_i} = n$, and $r_p = m$.
For the $6 \times 6$ Ferrers diagram $\CF =\{2,2,4,4,6,6\}$, we have $p=3$,  and for all $1 \leq i \leq 3$, ${\uu}_i = 2$ and $r_i = 2i$. This leads us to consider the first step of generalizing the Ferrers diagram $\CF =\{2,2,4,4,6,6\}$ as follows.

Let ${\uu},p$ be integers such that $m = {\uu} p$. Then consider the $m \times m$ Ferrers diagram 
\begin{equation}\label{F1}
\CF_1 = \{\overbrace{{\uu},\ldots,{\uu}}^{{\uu}\text{ times }},\overbrace{2{\uu},\ldots,2{\uu}}^{{\uu}\text{ times }},\dots,\overbrace{p{\uu},\ldots,p{\uu}}^{{\uu}\text{ times }}\}
\end{equation}

 $\CF_1$ can be further generalized as follows. 
For a positive integer $p$ consider a $p$-tuple of positive integers $(\uu_1, \ldots, \uu_p)$ and let $\uu = \max_{1\leq i \leq p} \uu_i$. Now consider the $m \times n$ Ferrers diagram 
\begin{equation}\label{F2}
\CF_2 = \{\overbrace{r_1,\ldots,r_1}^{{\uu}_1\text{ times }},\overbrace{r_2,\ldots,r_2}^{{\uu}_2\text{ times }},\dots,\overbrace{r_p,\ldots,r_p}^{{\uu}_p\text{ times }}\},
\end{equation}
 where  $r_i = k_i \uu$ for $1\leq i \leq p$ for any strictly increasing sequence of $p$ positive integers ${k_1, \ldots, k_p}$. 
 
% of values at most $q-1$. Note that the last condition on $k_i$'s implies $p < q$. 
It is clear that $\CF_1$ is a particular case of $\CF_2$. We construct $m \times n$ FDRM codes with minimum distance $d \leq \min\{m,n\}$ for $\CF_1$ and $\CF_2$ of dimension $v_0$. Recall that for a given $m \times n$ Ferrers diagram $\CF$ and an integer $d$ such that $1 \leq d \leq \min\{m,n\}$, the value of $v_0$ is the number of dots remaining after removing the rightmost $d-1$ columns from $\CF$. Now Theorem \ref{THM27} implies that $v_{\min}{[\CF,d]} \leq v_0$ and therefore the codes we construct are optimal FDRM codes. 

\textbf{Notation:} For the rest of this section we use $(\F[x])_a$ to denote the space of polynomials over $\F$ of degree at most $a-1$.

The following result provides construction of optimal FDRM codes for Ferrers diagrams of the form $\CF_2$ as given in \eqref{F2}.

\begin{theorem}\label{thm:37}
Let $(\uu_1, \ldots, \uu_p)$ be a $p$-tuple of positive integers with $p \geq 1$ and set $\uu = \max_{1\leq i \leq p} \uu_i$. Now for any set of $p$ positive integers $\{k_1, \ldots, k_p\}$ such that $k_1 < k_2 < \cdots <k_p$, consider the $m \times n$ Ferrers diagram 
\begin{equation}
\CF_2 = \{\overbrace{r_1,\ldots,r_1}^{{\uu}_1\text{ times }},\overbrace{r_2,\ldots,r_2}^{{\uu}_2\text{ times }},\dots,\overbrace{r_p,\ldots,r_p}^{{\uu}_p\text{ times }}\},
\end{equation}
 where  $r_i = k_i \uu$ for $1\leq i \leq p$, $m = r_p$ and $n = \sum_{i=1}^{p}\uu_i$. Then for any $d$ with $1 \leq d \leq n$, there exists an $m \times n$ optimal $[\CF_2,d]_q$ code for any prime power $q > k_p$.
 
\end{theorem}

\begin{proof}
First we calculate $v_0$ for the Ferrers diagram $\CF_2$ which is our target dimension to achieve. For a fixed $d$ such that $1 \leq d \leq n = \sum_{i=1}^{p}\uu_i$, there exist unique integers $t$ and $I$ such that $d-1 = \sum_{j=I}^p \uu_j -t$ where $0 \leq t < \uu_{I}$ and $1 \leq I \leq p$. So the number of dots in the first $n -d +1 = \sum_{i=1}^{I-1}\uu_i+t$ columns of $\CF_2$ is $v_0 = \sum_{i=1}^{I-1} r_i \uu_i + t r_{I}$.

%First we calculate the upper bound $v_{min}[\CF, d]$.
% We group the columns of the diagram $\CF$ into $n$ blocks $B_i$ for $1 \leq i \leq n$ where $B_i$ consists of the $m_i$ columns with $r_i$ dots in each column. Assume that the $(d-1)$-th column from the right is contained in the block $B_I$ for some integer $I$ with $1 \leq I \leq n$. 
%  Then $d-1 = \sum_{j=I}^n m_j - t$ for some integer $1\leq t < m_I$ and $M-d+1 = \sum_{i=0}^{I-1}m_i+t$.
%  So for any $\sum_{j=1}^{J-1}m_j < i \leq \sum_{j=1}^{J}m_j$, the $i$-th column have $ r_i = r_{J}$ many dots. Now $r_J = k_Jm \geq Jm \geq \sum_{j=1}^{J}m_j$ which implies $r_i \geq i$. Therefore applying Lemma \ref{lem:5} we get $ v_{min}[\CF,d] = v_0$. In other words, $v_{min}[\CF,d]$ is the number of dots after removing the last $d-1$ columns. So 
%\begin{equation}\label{eq:4} 
%v_{min}[\CF,d] = \sum_{j=1}^{I-1} r_jm_j + r_I t.
%\end{equation}
%For a fixed $\l \in \F_{q^\uu}$, let us denote by $\phi$ the automorphism $\phi_{q, \l}$ as defined in .
We consider the MRD code $\C_{\l,k}(\B)$ over $\F_{q^\uu}(x)/\K$ with $k=n-d+1$ and 
$ \B = \{a_1 x^{k_1-1},\dots,a_{\uu_1} x^{k_1-1}, a_1 x^{k_2-1},\dots,a_{\uu_2} x^{k_2-1},\ldots, a_1 x^{k_2-1},\dots,a_{\uu_p} x^{k_p-1}\}
$ with $\{a_1,\ldots,a_{\mu_i}\} \subseteq \{a_1,\ldots,a_{\mu}\}$ for all $1 \leq i \leq p$, where $\{a_1,\ldots,a_{\mu}\}$ is an ordered $\Fq$-basis of $\F_{q^\uu}$. Note that $\B$ is indeed a linearly independent set over the fixed field $\K$ of the corresponding automorphism $\fql$ since $q > k_i$ for all $1 \le i \le p$.

%Suppose that $\Fqm/\Fq$ is a finite field extension of degree $m$ with basis $\{ a_1,\dots,a_m\}$. 
%
%
%Suppose that $\l\in \Fqm^*$ such that $N(\l)$ is of order $q-1$ in $\Fq^*$. Define the automorphism $\f=\fql$ on $\Fqm(x)$ and let $\K$ be the fixed field.
%We choose$\K$-linearly independent elements of $\Fqm(x)$.

%Notice that because of the choice of $q$ to be larger than the $k_i$'s, these elements are linearly independent over the fixed field of $\phi_{q, \l}$.

%We evaluate the linear operators in $\Lf_k$ on the previous linearly independent elements, where $k = M-d+1 = \sum_{i=1}^{I-1}m_i + t$. We get a linear rank metric code with generator matrix $\GG$ over $\Fqm(x)$.

We consider the following generator matrix $\GG$ of $\C_{\l,k}(\B)$ in block form as follows.
\[
\GG = 
\begin{pmatrix}
\AA_{1,1}x^{k_1-1}   & \AA_{1,2}x^{k_2-1}   & \hdots & \AA_{1,{i}}x^{k_i-1}   & \hdots & \AA_{1,p}x^{k_p-1}   \\
\AA_{2,1}x^{k_1-1}   & \AA_{2,2}x^{k_2-1}   & \hdots & \AA_{2,i}x^{k_i-1}   & \hdots & \AA_{2,p} x^{k_p-1}   \\ 
\vdots      &     \vdots   & \ddots & \vdots      & \ddots & \vdots               \\
\AA_{I-1,1}x^{k_1-1} & \AA_{I-1,2}x^{k_2-1} & \hdots & \AA_{I-1,i}x^{k_i-1} & \hdots & \AA_{I-1,p} x^{k_p-1} \\ 
\AA_{I,1}x^{k_1-1}   & \AA_{I,2}x^{k_2-1}   & \hdots & \AA_{I,i}x^{k_i-1}   & \hdots & \AA_{I,p} x^{k_p-1}    
\end{pmatrix},
\]
where $\AA_{i,j} \in \F_{q^\uu}^{\uu_j\times \uu_j}$ and $\AA_{I,j} \in \F_{q^\uu}^{t\times \uu_{I}}$ for $1\leq i \leq I-1$, $1 \leq  j \leq p$.
 
Notice that the $\AA_{i,j}$'s are Moore matrices left multiplied with a diagonal matrix which has nonzero entries in the diagonal. Hence from Proposition \ref{PRO15} it follows that $\AA_{i,j}$ is invertible for every $1\leq i,j \leq I-1$. Therefore the generator matrix can be transformed into the following matrix.
%\[
%\GG' = 
%\begin{pmatrix}
%\AA_{1,1}x^{k_1-1}   & \AA_{1,2}x^{k_2-1}   & \hdots & \AA_{1,I}x^{k_i-1}   & \hdots & \AA_{1,n}x^{k_n-1}   \\
%\0   & \AA'_{2,2}x^{k_2-1}   & \hdots & \AA'_{2,I}x^{k_i-1}   & \hdots & \AA'_{2,n} x^{k_n-1}   \\ 
%\vdots      &     \ddots   & \ddots & \vdots      & \ddots & \vdots               \\
%\0 & \0 & \ddots & \AA'_{I-1,I}x^{k_i-1} & \hdots & \AA'_{I-1,n} x^{k_n-1} \\ 
%\0   & \0   & \hdots & \AA'_{I,I}x^{k_i-1}   & \hdots & \AA'_{I,n} x^{k_n-1}    
%\end{pmatrix}.
%\]
%
% which is further reduced by dividing some rows by an appropriate power of $x$ to get a much simplified the form.
 \begin{equation}\label{eq:5}
\GG' = 
\begin{pmatrix}
\AA_{1,1}   & \AA_{1,2}x^{k_2-k_1}   & \hdots & \AA_{1,I}x^{k_I-k_1}   & \hdots & \AA_{1,p}x^{k_p-k_1}   \\
\0   & \AA'_{2,2}   & \hdots & \AA'_{2,I}x^{k_I-k_2}   & \hdots & \AA'_{2,p} x^{k_p-k_2}   \\ 
\vdots      &     \ddots   & \ddots & \vdots      & \ddots & \vdots               \\
\0 & \0 & \ddots & \AA'_{I-1,I}x^{k_I-k_{I-1}} & \hdots & \AA'_{I-1,p} x^{k_p-k_{I-1}} \\ 
\0   & \0   & \hdots & \AA'_{I,I}   & \hdots & \AA'_{I,p} x^{k_p-k_I}    
\end{pmatrix}.
\end{equation}

Now let $V$ be the $\F_{q^\uu}$-subspace of $\F_{q^\uu}[x]^{n-d+1}$ defined by,
\begin{align*}
V =   \{ (c_{1,1},\dots,c_{1,{\uu}_1},c_{2,1},\dots&,c_{2,{\uu}_2},\ldots,c_{I-1,1},\dots,c_{I-1,{\uu}_{I-1}},c_{I,1} \ldots,c_{I,t})\colon   \\
& c_{i,j} \in \F_{q^\uu}[x] \text{ with } \deg c_{i,j} \leq k_i -1 \}.
\end{align*}
Note that 
\begin{align*}
\dim_{\Fq}V &= \sum_{i=1}^{I-1} \uu  k_i \uu_i +  \uu  k_{I} t \\
&= \sum_{i=1}^{I-1}   r_i {\uu}_i +  r_{I} t,
\end{align*}
which is equal to $v_0$.

Let $\mathbf{\alpha} = \{a_1, \dots, a_{\uu}, a_1x, \dots, a_{\uu}x, \dots, a_1x^{k_p-1}, \dots, a_{\uu}x^{k_p-1}\}$ be a basis of $(\F_{q^\uu}[x])_{k_p}/\Fq$.
Now we define
$$\C_{\CF_2} := \psi_{\alpha}(V\GG'),$$ where $\psi_{\alpha}$ is the $\Fq$-isometry between $(\F_{q^\uu}[x])_{k_p}^n$ and $\Fq^{k_p\uu \times n}$ by the means of coordinate matrices with respect to the basis $\mathbf{\alpha}$ (similar to the map in equation \eqref{eq:psi}).

We show that $\C_{\CF_2}$ defines an optimal $[\CF_2,d]_q$ rank-metric code with the assumption that $q>k_p$. Since $\psi_{\alpha}$ is an $\Fq$-isometry and $G'$ is injective, $\dim_{\Fq} \C_{\CF_2} = \dim_{\Fq} V = v_0$.
 As $VG' \subseteq \C_{\l,k}(\B)$ and the later is an MRD code of minimum rank distance $d$, it is clear that any codeword of $\C_{\CF_2}$ has rank at least $d$. Suppose minimum rank distance of $\C_{\CF_2}$ is $d'$ such that $d' >d$. In that case, $v_0(\CF_2,d') < v_0(\CF_2,d) =  v_0=\dim \C_{\CF_2}$, but this contradicts with the bound of $\dim \C_{\CF_2}$ in Theorem \ref{THM27}. Therefore, the minimum rank distance of $\dim \C_{\CF_2}$ should be $d$.

 What remains is to check that the codewords of $\C_{\CF_2}$ have the shape $\CF_2$. For some $v\in V$, the first ${\uu}_1$ coordinates of $vG'$ are polynomials in $(\F_{q^\uu}[x])_{k_1}$, i.e., polynomials of degree at most $k_1 -1$ and the next ${\uu}_2$ coordinates are polynomials in $(\F_{q^\uu}[x])_{k_2}$, i.e., polynomials of degree at most $k_2 -1$ and so on. Thus in the first ${\uu}_1$ columns of $\psi_{\alpha}(vG')$, entries in all but the first $k_1{\uu}$ rows are guaranteed to be zero. 
 %Writing the entries of $\xx$ as an $\Fq$-linear combination of  
%\[
%\{a_1, \dots, a_{\uu}, a_1x, \dots, a_{\uu}x, \dots, a_1x^{k_p-1}, \dots, a_{\uu}x^{k_p-1}\},
%\]
%we extract the coefficients in $\Fq$ to transform $\xx$ into a matrix $\MM_\xx$. Since the $\xx$ are of degree $k_1-1$ at most, so all but the first $k_1{\uu}$ rows in the corresponding columns of $\MM_\xx$ have zero entries.
Similarly in the next ${\uu}_2$ columns of $\psi_{\alpha}(vG')$, entries in all but the first $k_2{\uu}$ rows  are guaranteed to be zero and so on. Thus it is clear that the codewords have shape $\CF_2$. Hence it completes the proof that $\C_{\CF_2}$ is an optimal $[\CF_2,d]_q$ code for any $q>k_p$. 
\end{proof}

\begin{remark}\label{rem:5}
	Notice that we have to use the condition $k_p\leq q-1$ for the construction to work. More precisely, in order to get the necessary length of the code, $q$ must be large enough to get the extension $\F_{q^\uu}(x)/\K$ of desired degree.
 \end{remark}
%The previous theorem implies the following corollary. The FDRM codes in the corollary are generalizations of the Codes from the diagram in Figure \ref{fig::1}.

In the example below we apply the above method to construct FDRM codes $\C$ for the Ferrers diagram $\CF = \{2,2,4,4,6,6\}$ and minimum rank distance $4$. 

\begin{example}\label{exa:40}
Let $\F_{4^2}(x)$ be the rational function field in one variable over a finite field of size $16$. 
Let $\l$ be a primitive element of $\F_{4^2}$ and thus the norm $N(\l)$ over the extension $\F_{4^2}/\F_4$ is equal to $3$.
Denote the fixed field of the automorphism $\phi_{4,\l}$ on $\F_{4^2}(x)$ by $\K$. If $\{ a_1,a_2\}$ is a basis of $\F_{4^2}/\F_4$, then a basis of $\F_{4^2}(x)/\K$ is given by 
\[
\B = \{ a_1,a_2,a_1x,a_2x,a_1x^2,a_2x^2\}.
\]
Consider the Ferrers diagram $\CF = \{2,2,4,4,6,6\}$.

We want to construct an FDRM code $\C$ over $\F_4$ defined on the Ferrers diagram $\CF$ and with minimum distance $d=4$. By Theorem \ref{THM27}, $\dim_{\Fq} \C\leq 8$, i.e. the number of dots after removing the last three columns. 

Consider the MRD rank  metric code $\C_1\subset \F_{4^2}(x)^6$  constructed by evaluating linear operators defined by $\phi_{4,\l}$ and of degree $2$ at  most. A generator matrix of $\C_1$ is given by 

\[
\AA = \begin{pmatrix}
a_1 & a_2 & a_1x & a_2x & a_1x^2 & a_2x^2 \\
a_1^q & a_2^q & a_1^q\l x & a_2^q\l x & a_1^q\l^2x^2 & a_2^q\l^2x^2 \\
a_1 & a_2 & a_1\l^{1+q} x & a_2\l^{1+q} x & a_1\l^{2(1+q)}x^2 & a_2\l^{2(1+q)}.
\end{pmatrix}
\]

The standard generator matrix of $\C_1$ has the form
\[
\GG = \begin{pmatrix}
1 & 0 & 0 & r_1x & r_2x^2 & r_3x^2 \\
0 & 1 & 0 & s_1x & s_2x^2 & s_3x^2 \\
0 & 0 & 1 & t_1 & t_2x & t_3x
\end{pmatrix},
\]
where $r_i,s_i,t_i \in \F_{4^2}$.
Notice that the minimum distance of $\C_1$ is $6-3+1=4$. 
Now consider the $\Fq$-linear subcode $\C_2$ of $\C_1$ defined by $\C_2= V\GG$, where

\[
V =
\lbrace (c_1,c_2,c_3+c_4 x) 
\colon c_i\in \F_{4^2}
\rbrace
\]

Since the minimum distance of $C_1$ is equal to $4$ and $\C_2$ has a codeword of rank $4$, then the minimum distance of $\C_2$ is equal to $4$. Now, $\dim_{\Fq} \C_2 = \dim_{\Fq} V = 8$.

All codewords of $\C_2$ have the forms
\[
(c_1,c_2,c_3+c_4x,c_5+c_6x,c_7+c_8x+c_9x^2, c_{10}+c_{11}x+c_{12}x^2),
\]
where $c_i\in \F_{4^2}$. Hence we can expand $\C_2$ into an FDRM code $\C$ of dimension $8$ and minimum distance $4$.
\end{example}

\begin{remark}
The construction in \cite[Example III.16]{AG19} provides optimal codes over $\F_2$ with $d=4$. Our construction can be used with codes over $\F_q$ where $q \geq 4$ for any minimum distance $d$. Furthermore, we are providing the construction for a more general type of Ferrers diagrams.
\end{remark}

Next we show that we can use a method similar to the previous construction to obtain optimal FDRM codes for Ferrers diagrams which generalizes $\CF_2$. 
\begin{theorem}\label{pro:42}
Take a $p$-tuple of positive integers $(\uu_1, \ldots, \uu_p)$ and let $\uu = \max_{1\leq i \leq p} \uu_i$. For any strictly increasing sequence of $p$ positive integers ${k_1, \ldots, k_p}$, set $r_i = k_i \uu$ for all $1 \leq i \leq p$. Let $m,n$ be integers such that $m = r_p$ and $n = \sum_{i=1}^{p} \uu_i$. For a fixed integer $d$ such that $1 \leq d \leq n$, let $I, t$ be unique integers so that $n - d+1 = \sum_{i=1}^{I-1} \uu_i + t$ with $0 \leq I \leq p$ and $0 \leq t < {\uu}_{I}$.
 Now consider the following Ferrers diagram:
	\begin{align*}
	\CF_3 = \{\overbrace{s_{1,1},\ldots,s_{1,\uu_1}}^{\uu_1}, \,\dots, & \overbrace{s_{I-1,1},\ldots,s_{I-1,\uu_{I-1}}}^{\uu_{I-1}},\,
	\overbrace{s_{I,1},\ldots,s_{I,t}}^{t},
	\overbrace{r_{I},\ldots,r_{I}}^{\uu_{I}-t\text{ times }},\\ 
	&\overbrace{r_{I+1},\ldots,r_{I+1}}^{\uu_{I+1}\text{ times }},\dots,\overbrace{r_p,\ldots,r_p}^{\uu_p\text{ times }}\},
	\end{align*}
	where $s_{i,j} \leq r_i \text{ for } 1 \leq j \leq \uu_{i} \text{ and } 1 \leq i \leq I-1$ and $s_{I, j} \leq r_{I}$ for $1 \leq j \leq t$. Then there exists an optimal $[\CF_3,d]_q$ code for all $q > k_p$. Here $v_{min}[\CF_3,d] = v_0 $.
\end{theorem}

\begin{proof}
We want to construct an $[\CF_3,d]_q$ code with dimension $v_0$ which, in this case, is equal to $\sum_{i=1}^{I-1}(\sum_{j=1}^{\uu_i}s_{i,j}) + \sum_{j=1}^{t}s_{I,j}$. 
  
We consider the MRD code $\C_{\l,k}(\B)$ over $\F_{q^\uu}(x)/\K$ with $k=n-d+1= \sum_{i=1}^{I-1}\uu_i + t$.
Notice that the generator matrix $G'$ in Equation \eqref{eq:5} can be transformed into a block matrix such that $\AA_{i,j}$ for $1 \leq i,j \leq I$ are as follows.
	\[
	\begin{cases}
	   \AA_{i,i}\, = \II_{\uu_i}, & 1 \leq i \leq I-1,\\
	   \AA_{I,I} = [\II_t|*], &\\
	   \AA_{i,j}\, = \0,        & 1 \leq i,j \leq I-1,\; i \neq j,\\  
	   \AA_{i,I}\, = [\0_t|*],  & 1 \leq i \leq I.
	\end{cases}
	\]
	In other words, we consider the generator matrix $G''$ of $\C_{\l,k}(\B)$ of the form $[I_k | *]$ where $k = \sum_{i=1}^{I-1}\uu_i + t$.
	
%	Note that $n-d+1 = \sum_{i=1}^{I-1}\uu_i + t$. 
%	
%From Lemma \ref{lem:5}, for any Ferrers diagram $\CF_1 = \{c_1, \ldots, c_n\}$ such that $c_i = r_i$ for all $(M-d+2)\leq i \leq M$, the bound $v_{min}[\CF_1,d] = v_0$.	
%	 So we can construct optimal Ferrers diagram codes for even more general form 
%	\begin{align*}
%	\CF = \{\overbrace{s_{1,1},\ldots,s_{1,{\uu}_1}}^{{\uu}_{1}}, \,\dots, & \overbrace{s_{I-1,1},\ldots,s_{I-1,{\uu}_{I-1}}}^{{\uu}_{I-1}},\,
%	\overbrace{s_{I,1},\ldots,s_{I,t}}^{t},
%	\overbrace{r_{I},\ldots,r_{I}}^{{\uu}_{I}-t\text{ times }},\\ 
%	&\overbrace{r_{I+1},\ldots,r_{I+1}}^{{\uu}_{I+1}\text{ times }},\dots,\overbrace{r_n,\ldots,r_n}^{{\uu}_n\text{ times }}\},
%	\end{align*}
%	where $s_{i,j} \leq r_i \text{ for } 1 \leq i \leq I, 1 \leq j \leq {\uu}_{i}$.
 %Now  consider an $\Fq$-subspace $U_{1,1}$ of $\F_{q^\uu}[x]_{k_1}$ of dimension $s_{1,1}$. Let $\alpha_1$ is a $\Fq$-basis of 
%$U_{1,1}$. We extend this basis to get a $\Fq$-subspace of dimension $s_{1,2}$ of $\F_{q^\uu}[x]_{k_1}$. In this way we can take  $\Fq$-subspaces $U_{1,j}$ of dimension $s_{1,j}$ of $\F_{q^\uu}[x]_{k_1}$. Similarly we take $U_{i,j}$ to be $\Fq$-subspace of $\F_{q^\uu}[x]_{k_i}$ of dimension $s_{i,j}$ for $1 \leq i \leq I-1$ and $1 \leq j \leq {\uu}_i$. Continuing further, for $1 \leq j \leq t$, let $U_{I,j}$ be  

We construct a flag of $\Fq$-spaces $$ U_{1,1} \subseteq \cdots \subseteq U_{1,\uu_1} \subseteq \cdots \subseteq U_{I-1,1} \subseteq \cdots \subseteq U_{I-1, \uu_{I-1}}\subseteq U_{I,1} \subseteq \cdots \subseteq U_{I,t}$$ such that $U_{i,j} \subseteq (\F_{q^\uu}[x])_{k_i}$ of dimension $s_{i,j}$ for $1 \leq i \leq I-1$ and $1 \leq j \leq {\uu}_i$ and $U_{I,l} \subseteq (\F_{q^\uu}[x])_{k_I}$ for $1 \leq l \leq t$. We also take $\alpha_{i,j}$'s to be ordered $\Fq$-bases of $U_{i,j}$'s such that $\alpha_{i_1,j_1} \subseteq \alpha_{i_2,j_2}$ if $i_1 < i_2$ or if $i_1 = i_2$ and $j_1 < j_2$. In other words, the ordered bases $\alpha_{i,j}$'s also form a flag with respect to inclusion. Then we extend $\alpha_{I,t}$ to a $\Fq$-basis $\alpha_{I}$ of $(\F_{q^\uu}[x])_{k_I}$. Then for $I+1 \leq i \leq p$, we consider $\alpha_i$ to be $\Fq$-bases of $(\F_{q^\uu}[x])_{k_i}$ such that $\alpha_{i-1} \subseteq \alpha_i$. We set $\alpha = \alpha_{p}$, the ordered basis of $(\F_{q^\uu}[x])_{k_p}$ over $\Fq$.

Consider the following subspace of $\F_{q^\uu}[x]^{n-d+1}$:
\begin{align*}
V = \{(a_{1,1} ,\ldots, a_{1,{\uu}_1},\ldots,a_{I-1,1} ,\ldots, a_{I-1,{\uu}_{I-1}}, a_{I,1} ,\ldots, a_{I,t})\colon a_{i,j} \in U_{i,j} \}.
\end{align*}

We define $\C_{\CF_3}:= \psi_{\alpha} (VG'')$, where $\psi_{\alpha}$ gives the matrix representation of elements of $VG''$ with respect to the $\Fq$-basis $\alpha$ of $(\F_{q^\uu}[x])_{k_p}$.

It is clear to see that $\dim_{\Fq} V = \sum_{i=1}^{I-1}(\sum_{j=1}^{\uu_i}s_{i,j}) + \sum_{j=1}^{t}s_{I,j}$. Since $\psi_{\alpha}$ is an $\Fq$-isometry and $G''$ is injective, we have $\dim \C_{\CF_3} = \dim_{\Fq} V = v_0$. That the minimum rank distance of $\CF_3$ is $d$ and all the codewords of $\C_{\CF_3}$ have shape $\CF_3$ follow by the same arguments as given in the proof of Theorem \ref{thm:37}.
\end{proof}

\begin{example}
Let $\F_{4^3}(x)$ be the rational function field in one variable over a finite field of size $64$. Let $\l$ be a primitive element of $\F_{4^3}$ and thus the norm $N(\l)$ over the extension $\F_{4^3}/\F_4$ is equal to $3$.
Denote the fixed field of the automorphism $\phi_{4,\l}$ on $\F_{4^3}(x)$. If $\{ a_1,a_2,a_3 \}$ is a basis of $\F_{4^3}/\F_4$, then a basis of $\F_{4^3}(x)/\F_4$ is given by 
\[
\B = \{ a_1,a_2,a_3,a_1x,a_2x,a_3x,a_1x^2,a_2x^2,a_3x^2 \}.
\]
Consider the Ferrers diagram $\CF = \{1,2,3,5,5,9,9,9,9\}$.

We want to construct an FDRM code $\C$ over $\F_4$ defined on the Ferrers diagram $\CF$ and with minimum distance $d=5$. By Theorem \ref{THM27}, $\dim_{\Fq} \C\leq 16$, i.e. the number of dots after removing the last four columns. 

Consider the MRD rank  metric code $\C_1\subset \F_{4^3}(x)^9$  constructed by evaluating linear operators defined by $\phi_{4,\l}$ and of degree $4$ at  most. So considering the $(5 \times 9)$- generator matrix of $\C_1$ which have images of the elements in $\B$ as rows, can be transformed in the following standard form: 

%\comment{The matrix is still too big}
%{\small
%\[
%\GG_1 = \begin{pmatrix}
%a_1 & \vdots & a_3 & a_1x & \vdots & a_3x & a_1x^2 & \vdots & a_3x^2 \\
%a_1^q & \vdots & a_3^q & a_1^q\l x & \vdots & a_3^q\l x & a_1^q\l^2x^2 & \vdots & a_3^q\l^2x^2 \\
%a_1^{q^2} & \vdots & a_3^{q^2} & a_1^{q^2}\l^{1+q} x & \vdots & a_3^{q^2}\l^{1+q} x & a_1^{q^2}\l^{2(1+q)}x^2 & \vdots & a_3^{q^2}\l^{2(1+q)}x^2 \\
%a_1 & \vdots & a_3 & a_1\l^{1+q+q^2} x & \vdots & a_3\l^{1+q+q^2} x & a_1\l^{2(1+q+q^2)}x^2 & \vdots & a_3\l^{2(1+q+q^2)}x^2 \\
%a_1^{q^2} & \vdots & a_3^{q^2} & a_1^{q^2}\l^{1+q+q^2} x & \vdots & a_3^{q^2}\l^{1+q+q^2+q^3} x & a_1^{q^2}\l^{2(1+q+q^2+q^3)}x^2 & \vdots & a_3^{q^2}\l^{2(1+q+q^2+q^3)}x^2 \\
%\end{pmatrix}
%\]
%}

\[
\GG = \begin{pmatrix}
1 & 0 & 0 & 0 & 0 & r_1x & r_2x^2 & r_3x^2 & r_4x^2 \\
0 & 1 & 0 & 0 & 0 & s_1x & s_2x^2 & s_3x^2 & s_4x^2 \\
0 & 0 & 1 & 0 & 0 & t_1x & t_2x^2 & t_3x^2 & t_4x^2 \\
0 & 0 & 0 & 1 & 0 & u_1 & u_2x & u_3x & u_4x \\
0 & 0 & 0 & 0 & 1 & v_1 & v_2x & v_3x & v_4x \\
\end{pmatrix},
\]
where $r_i,s_i,t_i,u_s,v_i \in \Fqm$. 
Notice that the minimum distance of $\C_1$ is $9-5+1=5$. 
Now consider the $\Fq$-linear subcode $\C_2$ of $\C_1$ define by $\C_2= V\GG$, where

\begin{align*}
V =
\lbrace (
& c_1a_1,c_2 a_1 + c_2 a_2, c_3 a_1+ c_4 a_2 + c_5 a_3,  \\
& c_1' + (c_6 + c_7 a_1)x, c_2' + (c_8 + c_9 a_1)x ) \\
& \colon c_i\in \Fq \text{ and } c_j'\in \Fqm, 1\leq i\leq 9, 1\leq j\leq 2
\rbrace
\end{align*}

Since the minimum distance of $C_1$ is equal to $5$ and $\C_2$ has a codeword of rank $5$, then the minimum distance of $\C_2$ is equal to $5$. Now, $\dim_{\Fq} \C_2 = \dim_{\Fq} V = 16$.

All codewords of $\C_2$ have the form,
\begin{align*}
(
& c_1a_1,c_2 a_1 + c_2 a_2, c_3 a_1+ c_4 a_2 + c_5 a_3,  c_1' + (c_6 + c_7 a_1)x, c_2' + (c_8 + c_9 a_1)x, \\
& p_1(x),p_2(x),p_3(x),p_4(x)),
\end{align*}
where $c_i\in \Fq, c_i'\in \Fqm$ and $p_i(x)$ have degree $2$ at most. Hence we can expand $\C_2$ into an FDRM code $\C$ of dimension $16$ and minimum distance $5$.
\end{example}

A survey on the known main constructions of optimal FDRM codes with $v_{min}[\CF,d] = v_0$ can be found in \cite[Section 2]{LCF19}.

%Before going to the proof, here we analyze whether the known constructions can be used to derive any of the Ferrers diagrams described in the statement of Construction \ref{thm:37}.
%1) Etzion and Silberstein established the existence of the optimal $[\CF,k,d]_q$ rank metric codes where an $m \times n (m \geq n)$ Ferrers diagram
%and each of its rightmost $d -1$ columns has at least $m$ dots. The proof is based on the
%use of $q$-cyclic MRD codes. A better result is provided in \cite{EGRWZ16} with a simple proof by
%means of shortening systematic MRD codes. See also Theorem 23 of \cite{GR17} and Corollary 3.3 \cite{AG19}.
%
%2)Corollary \cite{ES09}: For $d \in \{1,2\}$, there exist optimal codes when $k$ = sum of dots in the first $n-d+1$ columns.
%
%3) Assume $\CF$ is an $m\times n$ Ferrers diagram and $m \geq n$. Let
%$2 \leq d \leq n-1$. If each of the rightmost $d-1$ columns in $\CF$ has at least $n-1$ dots, then
%there exists an $[\CF, k, d]_q$ code for any prime power q, where $k = \min{m - n + 1,\lambda_0 } + i=1$
%When $ \lambda_0 \leq m - n + 1$, the resulting FDRM code is optimal.
%It is remarked in \cite{AG19} that the reduction technique is very powerful in deducing new optimal FDRM code $[\CF_1,d]$ from an existing FDRM code $[\CF,d]$ where $\CF$ and $\CF_1$ satisfies following conditions: $\CF_1 \subsetneq \CF$ with $\mid \CF \setminus \CF_1 \mid =1$ and $v_{min}[\CF_1,d] = v_{min}[\CF,d] -1$. So the reduction technique can be readily applied in our case only if $d-1 \leq m_n$. 
%

Here we give an example of a Ferrers diagram $\CF$ for which, to the best of our knowledge, optimal FDRM code can not be formed using any of the previously known constructions based on subcodes of MRD codes. Then we show that the Theorem \ref{pro:42} gives an optimal rank-metric code in the Ferrers diagram $\CF$.
\begin{example}\label{exa:44}
Let $\CF = (5,5,5,5,5, 9,9,9,9,15,15,15,20,20,25)$ and suppose $d =12$. So $\CF$ is a $25 \times 15$-Ferrers diagram. We follow the notations of the above Theorem \ref{pro:42} and take $r_i = 5i$ and $m_i = 5-i+1$ for all $1 \leq i \leq 5$. Therefore, the Theorem \ref{pro:42} gives an optimal Ferrers diagram code for $\CF$ over $\Fq$ where $q > 5$.
Now if we notice the number of dots in the rightmost $d-1 = 11$ columns, we can directly conclude that we cannot apply any of constructions given in \cite[Theorem 2]{ES09}, \cite[Theorems 3, 8]{EGRWZ16}, \cite[Theorem 3.6]{AG19}, \cite[Construction 3.5]{LCF19}, \cite[Theorem 3.6]{ZG19}. As the $(d-1)$-th column of $\CF$ contains $9$ dots, any of the hypotheses of the respective theorems are not satisfied. 

\end{example}

\begin{remark}
Since our FDRM codes are equivalent to subcodes of some MRD code over the rational functions and the later has a decoding algorithm as we mentioned in Remark \ref{REM5}, then our FDRM code has a decoding algorithm. The complexity of the decoding algorithm is quadratic w.r.t. the length of the code, when considering operations over the rational functions.  In order to know the exact complexity when considering operations over $\Fqm$, we need to know the degree of the polynomials involved in the decoding algorithms. This depends on the shape of the Ferrers diagram we are considering. 
\end{remark}

\section{Conclusion}\label{sec:7}
%In this paper%
We define a class of automorphisms $\fql$ on $\Fqm(x)$ to construct $\C_{\l,k}(\B)$, a particular class of generalized Gabidulin codes of \cite{ALR18} (which are MRD codes), over the extension $\Fqm(x)/\K$ where $\K$ is the fixed field of $\fql$. Then we derive two applications of this construction.
%%the field of rational functions over finite fields%%. 
%We use these automorphisms to construct maximum rank distance code over rational function fields. Thus we have been able to give a particular class of generalized Gabidulin codes over arbitrary fields as introduced in \cite{ALR18}. 

Firstly, we obtain MRD codes over finite fields by reducing the codes $\C_{\l,k}(\B)$ to finite fields. %to get MRD codes over finite fields.%
Furthermore, in this process we provide MRD codes which are not that equivalent to (generalized) Gabidulin codes and (generalized) twisted Gabidulin codes. In fact, we show that for such a code $\C$ over $\F_{q^m}/\Fq$, the intersection $\C \cap \C^{q^s}$ is trivial for any integer $s$ with $\gcd(m,s)=1$. This property is an advantage because cryptosystems using such codes are resistant to distinguisher attack based on a method of Overbeck. Though in our construction, the degree of the extension used is large compared to the length of the code, but this confirms the existence of non-Gabidulin codes when the base field is large enough as proved in \cite{NHTRR18}.  

 Our next construction is of optimal Ferrers diagram codes. Our codes include some Ferrers diagrams for which there was no known construction for optimal codes. But the construction works on large base fields $\Fq$. Thus constructing optimal FDRM codes over any base field $\Fq$ for the Ferrers diagrams we considered still remains open. 
 %question about the existence of optimal Ferrers diagram codes for the diagram that we are working with. In particular, it is still not known if Ferrers diagram codes over $\F_2$ exist or not.

As another application, we give a new construction of a known class of maximal sum rank distance (MSRD) codes, called linearized Reed-Solomon codes, of \cite{MK19}. We discuss the construction in the appendix since the code we obtain is not new.

As a concluding remark, we mention that we only considered a particular type of automorphisms for the construction of the codes $\C_{\l,k}(\B)$ over rational functions. So it will be interesting to study the rank-metric codes constructed via other automorphisms of rational function fields and investigate if we can find optimal rank-metric codes for new Ferrers diagrams for which no optimal construction is not known yet.

%{\color{green} CHeck the first reference. Do we need to change the date of access? }
%\begin{thebibliography}{AAAA}
	
%	\addtocontents{toc}{\hfill{}\\{\bf References} \hfill{\bf \thepage}\\}
%	\thispagestyle{empty}
%\end{thebibliography}
\bibliography{references}

\begin{appendices}
\section{Sum rank-metric codes}\label{sec:A}
%The type of network coding proposed in \cite{KK08-A} is called \emph{one-shot} network coding as the network channel is only used once. Another type of coding is the multishot network coding and the sum rank metric can be used to deal with this setting \cite{MBK15,MP19,MK19,MK18,NU10}. The codes embedded with this metric, called sum rank metric codes, can also be used in space time coding \cite{GH03} as well as in distributed storage \cite{MK18}. 

%In this section, we show how the new construction of MRD codes over rational function fields can be used to recover the \textit{linearized Reed-Solomon codes} given in \cite[Def. 11]{MK19}. These codes form a particular class of the general linearized Reed-Solomon codes defined in \cite[Def. 31]{MP18}. 

First we recall some basic definitions and facts about sum rank-metric codes before reconstructing the \textit{linearized Reed-Solomon codes} of \cite[Def. 11]{MK19}.

\begin{definition}[Sum rank metric]
Let $n_i$ be positive integers for $i =1, \ldots, l$ for some $l > 0$. For $\xx = (\xx_1|\xx_2|\dots|\xx_l)$ with $\xx_i\in \Fqm^{n_i}$, the sum rank of $\xx$ is defined as
\[
\sr (\xx) = \sum_{i=1}^l{\rank\,(\xx_i)},
\]
where $\rank \,(\xx_i)$ is the usual rank norm defined over the extension $\Fqm/\Fq$. It induces a metric, called sum rank metric, on $\Fqm^N$ where $N=\sum_{i=1}^l n_i$, as follows:
\begin{align*}
d_{sr}:\,\,  &(\Fqm^N)^2 \longrightarrow \N \\
          &(\xx,\yy)\,\, \mapsto \sr (\xx - \yy).
\end{align*}

\end{definition}

  A subspace $\C \subseteq \Fqm^{N}$ endowed with the sum rank metric is called a sum rank-metric code over $\Fqm$. The minimum sum rank distance of the linear code $\C$ is $d_{sr}(\C) = \min\{d_{sr}(\xx) : \xx \in \C \setminus \{0\}\}$. We call $\C  \subseteq \Fqm^{N}$ to be $[n_1,\dots, n_l;k,d]$-sum rank-metric code if it has minimum sum rank distance $d$ and dimension $k$.
  %For simplicity we omit the individual block lengths and call it as $[N,k,d]$-sum rank metric code.
For the rest of the discussion, we fix the partition of $N=\sum_{i=1}^{l}n_i$ and for brevity, denote an $[n_1,\dots, n_l;k,d]$-sum rank-metric code as $[N, k,d]$-sum rank-metric code.

The sum rank-metric codes satisfy the following Singleton bound.
\begin{theorem}[\cite{MP18,MK19}]
Let $\C$ be an $[N,k,d]$-sum rank-metric code. Then
\[
d\leq N-k+1.
\]
Codes attaining the bound are called maximum sum rank distance (MSRD) codes. 
\end{theorem}

Several MSRD codes have been constructed in \cite{MBK15,NPRV17,MP18,MK19}. We present a new way of constructing the MSRD codes, called linearized Reed-Solomon codes, defined in \cite[Def. 31]{MP18}, using the generalized Gabidulin codes over rational function fields. We recall some notations and the definition of linearized Reed-Solomon codes from \cite{MK19}.

% Their construction has a decoding algorithm which has complexity quadratic w.r.t. the length of the code. Moreover, this construction is just a particular case of a more general construction in \cite{MP18}.

Let $\B^{(i)} = \{\beta^{(i)}_1,\dots,\beta^{(i)}_{n_i}\} \subseteq \Fqm$ be linearly independent sets over $\Fq$ for $i=1,\dots,l$, and denote $\B = (\B^{(1)},\dots,\B^{(l)})$, where the $\B^{(i)}$'s can have common elements. 
%Let $\l$ be a primitive element of $\Fqm$.
For $\l$ in $\Fqm$, we define the operator $\D_\l$ on $\Fqm$ by 
\begin{align*}
\D_\l\colon \Fqm &\rightarrow \Fqm \\
a&\mapsto a^q \l.
\end{align*}
We take $\D_\l^1 = \D_\l$ and 
recursively define $\D_\l^i = \D_\l\circ \D_\l^{i-1}$. Thus $\D_\l^i(a)=a^{q^i} \l^{*i}$ for $a\in \Fqm$,
where $\l^{*i} = \l^{1+q+\dots+q^{i-1}}$. 

By abuse of notation, we also denote by $\D_\l$ the map on $\Fqm^{n_i}$, where we apply $\D_\l$ coordinatewise. 
%Now we are ready to describe the construction from \cite{MK19} in the following definition.

\begin{definition}\cite[Definition 11]{MK19}\label{MSRDUmberto}

Let $\l$ be a primitive element of $\Fqm$. The linearized Reed-Solomon sum rank-metric code $\C(\B,\l)$ over $\Fqm/\Fq$ is the $[N,k,N-k+1]$-linear code with generator matrix
\[
\GG_{(\B,\l)} = \begin{pmatrix}
\B^{(1)} & \B^{(2)} & \dots & \B^{(l)}\\
\D_1(\B^{(1)}) & \D_{\l}(\B^{(2)}) & \dots & \D_{\l^{l-1}}(\B^{(l)})\\
\vdots & \vdots & \ddots & \vdots \\
\D_1^{k-1}(\B^{(1)}) & \D{\l}^{k-1}(\B^{(2)}) & \dots & \D_{\l^{l-1}}^{k-1}(\B^{(l)})\\
\end{pmatrix}.
\]
\end{definition}

In \cite{MP18}, it is shown that the code $\C(\B,\l)$ is a maximum sum rank-metric code of length $N=\sum_{i=1}^l n_i$ and dimension $k$.

The following short will be useful for our construction of MSRD code of Definition 
\ref{MSRDUmberto}.
%, we prove a short lemma which play a role in the construction.  

\begin{lemma}\label{LEM49}
Let $N=\sum_{i=1}^l n_i$. Then for \[\cc = (c_{1,1}, \ldots, c_{1,n_1}| c_{2,1} , \ldots, c_{2,n_2}| \ldots| c_{n,1},\ldots, c_{n ,n_l}) \in \Fqm^N, \text{ we have }\] 
%be an element of $\Fqm^N$ where $N=\sum_{i=1}^l n_i$. 
 \[
 \sr (\cc) = \rank \, (c_{1,1}, \ldots, c_{1,n_1}, c_{2,1} x, \ldots, c_{2,n_2} x, \ldots, c_{n,1} x^{l-1},\ldots, c_{l,n_l} x^{l-1}),
 \]
  where on the left-hand side $\sr (\cc)$ is the sum rank norm, and on the right-hand side we have the usual rank norm for the extension $\Fqm(x)/\Fq$, i.e. the dimension of the $\Fq$-subspace of $\Fqm(x)$ generated by 
 
\[\{ c_{1,1}, \ldots, c_{1,n_1}, c_{2,1} x, \ldots, c_{2,n_2} x, \ldots, c_{l,1} x^{l-1},\ldots, c_{l,n_l} x^{l-1}\}.
\]
\end{lemma}
\begin{proof}
The Lemma follows from the fact that any entry $c_{i,j} x^{i-1}$ of $\cc$ is linearly independent over $\Fq$ with the set of entries of $\cc$ which have power of $x$ different from $i-1$. Therefore, the contribution of $\{ c_{i,1} x^{i-1}, \ldots, c_{i,n_i} x^{i-1}\}$ in $\sr\, (\cc)$ is equal to $\rank\,\, (c_{i,1}, \ldots,  c_{i,n_i})$.
\end{proof}
%Now we are ready to give a construction of sum rank metric codes. Then we show that the codes constructed are MSRD codes and in fact, they are same codes as given in Definition \ref{MSRDUmberto}.

Now we proceed to give a construction of sum rank-metric codes. 

%For that, we consider $l \leq q-1$ and the automorphism $\phi = \fql$ of $\Fqm[x]$ as defined in Section \ref{sec:3}, i.e. $\fql(\sum_{i=0}^k f_i x^i) = \sum_{i=0}^k f_i^q \l^i x^i$, where $\l \in \Fqm^\times$ is such that $N(\l)$ has order $q-1$ in the multiplicative group $\Fq^\times$. We use this automorphism to define sum rank metric codes as follows. 

\begin{definition} \label{def:50}
	Let $\{a_1, \ldots, a_m\}$ be a basis of $\Fqm/\Fq$ and let $l\leq q-1$. Consider 
 \[
 \B_1 = \{a_1, \ldots, a_{n_1} \mid a_1 x, \ldots, a_{n_2} x \mid \ldots \mid a_1 x^{l-1},\ldots, a_{n_l} x^{l-1}\},
 \] 
 where $N = \sum_{i=1}^{l}n_i$ and $m \geq n_i$. 
 Let $k = N-d+1$ for some fixed integer $d$. Let $\l$ be a primitive element of $\Fqm^{\times}$ and $\fql$ be the automorphism of $\Fqm(x)$ as defined in Eq. \eqref{auto}.
 Suppose $\GG$ is the $k \times n$ matrix whose $i$-th row is $\fql^{i-1}$ applied on $\B_1$.

 Taking $V = \Fqm^k$ and $\overline{\C}=V\GG$, we consider a map $\Phi$ from $\overline{\C}$ to $\Fqm^N$, where 
\begin{align*}
&\Phi(c_{1,1}, \ldots, c_{1,n_1}\mid c_{2,1}x , \ldots, c_{2,n_2}x \mid \ldots \mid c_{l,1}x^{l-1},\ldots, c_{l, n_l}x^{l-1}) \\
&= (c_{1,1}, \ldots, c_{1,n_1}\mid c_{2,1} , \ldots, c_{2,n_2}\mid \ldots\mid c_{l,1},\ldots, c_{l, n_l}).
\end{align*}
We define the sum rank-metric code $\C_1$ to be the image $\Phi(\overline{\C})$.

\end{definition} 
 Note that $\overline{\C}$ is a subset of the MRD code $\C_{\l,k}(\B_1)$ of length $N$ and dimension $k$ over the rational functions $\Fqm(x)$ generated by $\GG$. We obtain the code $\C_1$ by omitting the powers of $x$ in the entries of elements in $\overline{\C}$. Therefore, the codewords of the linear code $\C_1$ over $\Fqm/ \Fq$ are of the form \[\cc = (c_{1,1}, \ldots, c_{1,n_1}\mid c_{2,1} , \ldots, c_{2,n_2}\mid \ldots\mid c_{l,1},\ldots, c_{l, n_l,}),\] where $c_{ij} \in \Fqm$.

The following result proves that $\C_1$ attains the Singleton bound and it is same to the linearized Reed-Solomon code in Definition \ref{MSRDUmberto}.
 \begin{theorem}\label{thm:51}
 The code $\C_1$ in Definition \ref{def:50} is a maximum sum rank distance code. Moreover, if we take $\B = (\B^{(1)},\dots,\B^{(n)}) $ in Definition \ref{MSRDUmberto} with $\B^{(i)} = (a_1,\dots,a_{n_i})$, then $\C_1 = \C(\B,\l)$.
 \end{theorem}
 \begin{proof}  
 From injectivity of $\Phi$, it follows that $\dim_{\Fqm}\C_1 = k$. We know that $\C_{\l,k}(\B_1)$ has minimum rank distance $d = N-k+1$.
As $\overline{\C} \subseteq \C_{\l,k}(\B_1)$, Lemma \ref{LEM49} implies $d_{SR}(\C_1) \geq N-k+1$. So combining with the Singleton bound we get, $d_{SR}(\C_1) = N-k+1$ and thus $\C_1$ is indeed an MSRD code. 

From the definitions it is clear that the generator matrix $\GG_{(\B,\l)}$ as in Definition \ref{MSRDUmberto} is obtained by applying the map $\Phi$ on the rows of the generator matrix $\GG$ of $\overline{\C}$. Thus $\C_1 = \C(\B,\l)$.
\end{proof}

We chose $l\leq q-1$, as in order to have $l$ blocks, we need to have a power $x^{l-1}$ in the original code over $\Fqm(x)$.
Notice that if we choose the independent elements for evaluation to be 
 \[
 \B_1 = \{a_1, \dots, a_{n_1}\},
 \]
then we get the construction of Gabidulin codes. Similarly, if we consider the evaluation on
 \[
  \B_1 = \{1,x,\dots,x^{n-1}\},
 \] 
Then we get Reed-Solomon codes. Thus as mentioned in \cite{MK19}, the construction gives a generalization of both the Gabidulin and Reed-Solomon codes.
\end{appendices}

%\bibliography{references}
\end{document}